\documentclass{JFM-FLM_Au}

\usepackage{enumitem}

\usepackage{tikz}
\usetikzlibrary{arrows.meta}
\definecolor{newtblue}{RGB}{40,90,160}
\definecolor{newtblueB}{RGB}{90,150,210}
\definecolor{oddred}{RGB}{200,70,20}
\definecolor{drv}{RGB}{120,120,120}
\definecolor{gridcol}{RGB}{190,190,190}

\lefttitle{F. {Vega Reyes}}
\righttitle{Journal of Fluid Mechanics}

\title{Steady base states in a two-dimensional chiral fluid. The chiral Stokes cavity}

\author{Francisco Vega Reyes\aff{1}}

\affiliation{\aff{1}Departamento de F\'isica and Instituto de Computaci\'on Cient\'ifica Avanzada (ICCAEx),
  Universidad de Extremadura, Avda. Elvas s/n, 06071 Badajoz, Spain}

\corresau{F. Vega Reyes, \email{fvega@eaphysics.xyz}}

\begin{document}

\maketitle

\begin{abstract}
  We develop from first principles the hydrodynamics of a two-dimensional chiral fluid, i.e.\ one carrying a
  net microscopic angular-momentum (spin) field. Enforcing angular-momentum conservation without imposing
  stress-tensor symmetry, we derive the full form of the stress tensor and of the spin flux, and we show that
  the entire chiral response is generated from the classical Newtonian one by a single operation of direct
  physical origin --- the $90^{\circ}$ rotation through which chirality acts, mirrored at the particle level by
  transverse forces \citep{Caprini_2025}. Applied to the irreducible (deviatoric) decomposition, the rotation
  assigns to each classical channel a chiral partner --- pressure to chiral pressure, bulk and shear
  viscosities to their odd counterparts, the spin-flux gradient to its rotated image --- one coefficient and
  one mechanical action per channel, with no further cross-couplings. In this representation the steady base
  states become elementary. Quiescent states are organised by a holomorphic chiral complex potential, the
  mechanical and chiral pressures forming a conjugate harmonic pair subject to a topological existence
  condition; inhomogeneous activity forces azimuthal flows; and a boundary-driven confined flow, the
  \textit{chiral Stokes cavity}, obeys a modified Helmholtz--Poisson system, solved in closed form in a
  circular domain and numerically in a square one. A single dimensionless group controls both geometries and
  sets the crossover from a screened, single-vortex regime to a sequence of sign-reversing vortical
  structures. The theory yields quantitative predictions, amenable to direct comparison with experiments on
  air-fluidised chiral disks \citep{Lopez_Castano_2022}, and recovers the phenomenological frameworks of the
  chiral-fluid literature as particular cases.
\end{abstract}    



\section{Introduction}
\label{sec:intro}

\subsection{Background}
\label{sec:background}

Active matter has attracted sustained interest across physics, biology, chemistry and medicine, driven both by
fundamental questions and by a wide range of applications \citep{Marchetti_2013}. Active fluids --- composed
of self-propelled units with an internal source of motility --- are among the most studied active systems,
exhibiting phenomena such as phase separation \citep{Cates_2015}, crystallisation
\citep{Omar_2021,Digregorio_2018} and caging transitions \citep{Debets_2023} broadly analogous to those found
in granular fluids \citep{Aranson_2006}.  The present work concerns a special class of active fluids in which
chiral symmetry is broken at the particle level \citep{Bowick_2022,Caprini_2025}. Particle-scale chirality
generically induces collective dynamics for which parity and time-reversal symmetry are violated, and these
broken symmetries leave a direct imprint on the macroscopic hydrodynamics: the stress tensor of the fluid
acquires an antisymmetric component \citep{Avron_1998,Banerjee2017,Beppu2021,Lou_2022}.

\begin{figure}
\begin{center}
\begin{tikzpicture}[>=Stealth, line cap=round, line join=round,
    frm/.style={font=\footnotesize},
    lbl/.style={font=\small},
    sq/.style={draw=black, thick},
    trA/.style={->, newtblue,  line width=1.3pt},
    trB/.style={->, newtblueB, line width=1.3pt},
    trO/.style={->, oddred,    line width=1.3pt},
    spin/.style={->, drv, thick}]

\def\a{1.0}
\def\off{0.18}
 
\node[lbl] at (-2.1,1.85) {(a)};
\node[frm] at (0,2.05) {kind I (achiral)};
 
\begin{scope}
  \draw[sq] (-\a,-\a) rectangle (\a,\a);
 
  \draw[trA] (\a+\off,-0.55) -- (\a+\off,0.55);
  \draw[trA] (-\a-\off,0.55) -- (-\a-\off,-0.55);
  \node[frm, newtblue] at (\a+0.72,0.78) {$\sigma_{yx}$};
 
  \draw[trB] (-0.55,\a+\off) -- (0.55,\a+\off);
  \draw[trB] (0.55,-\a-\off) -- (-0.55,-\a-\off);
  \node[frm, newtblueB] at (1.0,\a+0.48) {$\sigma_{xy}$};
 
  \draw[spin] (-0.45, 0.45)+(0:0.13)   arc[start angle=0,   end angle=300, radius=0.13];
  \draw[spin] ( 0.45, 0.45)+(300:0.13) arc[start angle=300, end angle=0,   radius=0.13];
  \draw[spin] ( 0.00, 0.00)+(0:0.13)   arc[start angle=0,   end angle=300, radius=0.13];
  \draw[spin] (-0.45,-0.45)+(300:0.13) arc[start angle=300, end angle=0,   radius=0.13];
  \draw[spin] ( 0.45,-0.45)+(0:0.13)   arc[start angle=0,   end angle=300, radius=0.13];
 
  \draw[->, newtblue,  thick] (-0.62,-2.0)+(-60:0.20) arc[start angle=-60, end angle=240, radius=0.20];
  \node[frm] at (-0.10,-2.0) {$+$};
  \draw[->, newtblueB, thick] ( 0.42,-2.0)+(240:0.20) arc[start angle=240, end angle=-60, radius=0.20];
  \node[frm] at (1.05,-2.0) {$= 0$};
 
  \node[frm] at (0,-2.65) {$\delta^{2}\left(\sigma_{yx}-\sigma_{xy}\right) = 0$};
  \node[frm] at (0,-3.12) {$\Omega = 0 \;\Rightarrow\; \sigma_{xy} = \sigma_{yx}$};
\end{scope}

\node[lbl] at (3.7,1.85) {(b)};
\node[frm] at (5.8,2.05) {kind II (chiral)};
 
\begin{scope}[shift={(5.8,0)}]
  \draw[sq] (-\a,-\a) rectangle (\a,\a);
 
  \draw[trO] (-0.55,\a+\off) -- (0.55,\a+\off);
  \draw[trO] (\a+\off,0.55) -- (\a+\off,-0.55);
  \draw[trO] (0.55,-\a-\off) -- (-0.55,-\a-\off);
  \draw[trO] (-\a-\off,-0.55) -- (-\a-\off,0.55);
 
  \draw[spin] (-0.45, 0.45)+(0:0.13) arc[start angle=0, end angle=300, radius=0.13];
  \draw[spin] ( 0.45, 0.45)+(0:0.13) arc[start angle=0, end angle=300, radius=0.13];
  \draw[spin] ( 0.00, 0.00)+(0:0.13) arc[start angle=0, end angle=300, radius=0.13];
  \draw[spin] (-0.45,-0.45)+(0:0.13) arc[start angle=0, end angle=300, radius=0.13];
  \draw[spin] ( 0.45,-0.45)+(0:0.13) arc[start angle=0, end angle=300, radius=0.13];
 
  \draw[->, oddred, thick] (120:1.55) arc[start angle=120, end angle=-120, radius=1.55];
  \node[frm, oddred] at (2.28,0.42) {net couple};

  \node[frm, oddred] at (1.2,\a+0.48) {$\sigma_{xy}$};
  \node[frm, oddred] at (\a+0.72,0.78) {$\sigma_{yx}$};
  \node[frm, oddred] at (\a-2.0,-1.4) {$p_R$};

  \draw[->, oddred,  thick] (-0.84,-2.0)+(-60:0.20) arc[start angle=240, end angle=-60, radius=0.20];
  \node[frm] at (-0.10,-2.0) {$+$};
  \draw[->, oddred, thick] ( 0.42,-2.0)+(240:0.20) arc[start angle=240, end angle=-60, radius=0.20];
  \node[frm] at (1.05,-2.0) {$= 0$};

  \node[frm] at (0,-2.65) {$\delta^{2}\left(\sigma_{yx}-\sigma_{xy}\right) = -\,2\,p_R\,\delta^{2} \neq 0$};
  \node[frm] at (0,-3.12) {$\sigma^{\mathrm{anti}}_{ij} = p_R\,\epsilon_{ij}$};
\end{scope}
 
\end{tikzpicture}
\end{center}
\caption{Why a non-vanishing spin field forces an antisymmetric stress. (a) Achiral fluid: the tangential
tractions exerted by the neighbouring fluid on a square element form two couples of opposite sense --- that
of the $x$-faces ($\sigma_{yx}$) and that of the $y$-faces ($\sigma_{xy}$). With the particle spins
averaging to zero, the element has no internal reservoir to absorb a net torque (an uncancelled couple
would impart a divergent angular acceleration as the element shrinks), so the couples must cancel:
$\sigma_{xy}=\sigma_{yx}$. (b) Chiral fluid: coherent particle spin and continuous injection of angular
momentum provide such a reservoir; the two couples co-rotate and the antisymmetric stress
$p_R\,\epsilon_{ij}$ exerts the uncompensated couple $-2p_R$ per unit area --- precisely the exchange term
of the spin balance \eqref{eq:angular_momentum_convective_form}. Drawn for $p_R>0$.
\label{fig:stress_symmetry}}
\end{figure}
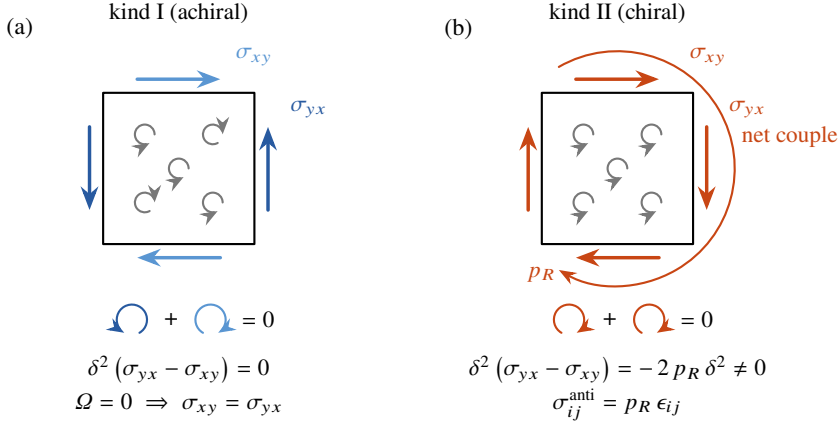

This departure from stress symmetry is not a technical detail but a foundational one. The symmetry of the
stress tensor in classical fluid mechanics is a direct consequence of the absence of a macroscopic angular
momentum field \citep{Batchelor_1967}: if every fluid element carries no net angular momentum, the torques
exerted by neighbouring elements must cancel, which forces the stress to be symmetric. In a chiral active
fluid this condition fails --- as a consequence of chirality, there is continuous injection of angular
momentum into the fluid --- and the Navier--Stokes framework built on a symmetric stress tensor is no longer
applicable; this argument is illustrated in figure~\ref{fig:stress_symmetry}. A distinct hydrodynamic theory,
with qualitatively new structure and flow phenomena, is required.

\par\medskip

\par\medskip

Existing treatments of the odd stress in chiral active fluids have largely proceeded either through
phenomenological symmetry arguments \citep{Avron_1998,Banerjee2017} or via Hamiltonian and Poisson-bracket
frameworks that impose stress symmetry at the outset \citep{Markovich2021}. A notable exception is the
linear-irreversible-thermodynamics treatment of \citet{Markovich_2019}, which retains the spin as a dynamical
variable and does not assume a symmetric stress; it is, however, formulated in three dimensions, and its
authors observe that the torque-dipole decompositions they classify cannot be realised in strictly
two-dimensional geometries, a case they leave aside. A systematic, first-principles derivation of the full odd
stress tensor --- of the kind long standard in classical fluid mechanics following \citet{Batchelor_1967} ---
has remained absent. In the present work we supply this derivation. Enforcing angular momentum conservation
without assuming stress symmetry yields an odd stress that decomposes naturally and without approximation into
two structurally distinct parts: an anisotropic, symmetric traceless contribution proportional to the
deviatoric strain rate, and an isotropic antisymmetric contribution that takes the form of a chiral pressure
$p_R \varepsilon_{ij}$. This decomposition is not merely a notational convenience: the two parts play entirely
different dynamical roles. The anisotropic part, being symmetric, makes no contribution to the vorticity
equation regardless of the flow. The chiral pressure term, by contrast, is the sole driver of the chiral
circulation. Using the deviatoric strain rate $d_{ij}$ rather than the full strain rate $e_{ij}$ --- as in
some earlier formulations \citep{Banerjee2017} --- renders this decomposition exact and free of spurious
compressibility and antisymmetric contributions, giving the result a structural cleanliness appropriate to a
Batchelor-style treatment. In our description, the dynamical spin field~$\Omega(\boldsymbol{r}, t)$ emerges
naturally from the competition between the applied torque, the rotational drag on the particles and the fluid
vorticity~$\omega(\boldsymbol{r}, t)$. We also derive the full form of the spin flux.

From our description, a modified Helmholtz--Poisson system governing confined chiral flows follows
directly. The key outcome of the derivation is a single dimensionless parameter $\alpha$ that encodes the
coupling between the particle spin and the fluid vorticity, and whose sign distinguishes active pumping from
damped relaxation. We then analyse in detail the steady vortical flow driven by a confining boundary --- a
problem structurally analogous to the classical lid-driven cavity flow, which we term the \textit{chiral
  Stokes cavity flow} --- solving it in closed form in a circular domain and numerically in a square one, and
compare its predictions quantitatively with experimental data obtained from chiral-disk suspensions
\citep{Lopez_Castano_2022}.

\subsection{Motivation and scope}
\label{sec:motivation}

The foundations of classical fluid mechanics follow an incremental logic that has proved extraordinarily
productive: one begins with incompressible flow at constant transport coefficients, identifies the governing
dimensionless groups, constructs the canonical steady solutions with high geometric symmetry, characterises
their transient relaxation, and then examines stability, instability and turbulence. Each layer of knowledge
provides the reference state for the next.

An analogous body of knowledge is now emerging for chiral active fluids
\citep{Avron_1998,Banerjee2017,Beppu2021,Bowick_2022}, yet a corresponding library of simple, analytically
tractable steady-flow solutions has not been established.  This gap is more than a matter of completeness:
without well-understood base states it is difficult to isolate the physical mechanisms that distinguish chiral
from non-chiral flows, and harder still to compare theory with experiment in a controlled and quantitative
way. The present paper is a step towards closing this gap.

The physical motivation is broad. Chiral interactions arise in a variety of biological and synthetic systems:
the $\mathrm{F}_1$ ring of ATP synthase produces a directional torque through its geometrically chiral
coupling to the molecular stalk \citep{Boyer1997,Noji1997}; bacterial ATP synthase can drive chiral edge
currents and vortices whose handedness is set by the coupling to flagellar motors \citep{Yoshida_2001};
kinesin \citep{Vale1985} and myosin generate directed motion through chiral lever arms and asymmetric ATP
hydrolysis cycles \citep{Sweeney2010}. An extended hydrodynamic theory for chiral active fluids is therefore
needed to describe important phenomena at the biological scale \citep{Beppu2021}.

Suspensions of self-propelled particles with broken mirror symmetry represent a class of active matter in
which microscopic chirality drives macroscopic coherent flows. The collective dynamics of swimming bacteria
exemplify this principle: dense bacterial suspensions exhibit turbulent-like spatiotemporal behaviour at
scales much larger than individual cells, with characteristic vortex structures and preferential rotational
directions determined by the inherent chirality of flagellar propulsion \citep{Sokolov2007}. Recent
experiments on confined bacterial suspensions reveal that this interplay between individual chiral swimming
and boundary conditions generates geometry-independent collective edge currents whose sense of rotation aligns
with the handedness of bacterial propulsion, thereby breaking the mirror symmetry expected from achiral
confinement geometries \citep{Beppu2021}. These observations underscore a fundamental question: how do the
torques generated by spinning particles couple to the vorticity field of the surrounding fluid, and what
continuum equations govern the resulting steady flows?

Beyond the active matter context, the mathematical structure derived here has notable analogues in
astrophysical fluid dynamics. Magnetised plasmas in differentially rotating discs --- galactic discs and
accretion flows alike --- naturally exhibit Hall viscosity, or gyroviscosity in the Braginskii sense
\citep{Braginskii1965}, which is the astrophysical counterpart of the odd stress tensor derived below. In both
settings the non-dissipative transverse stress arises from intrinsic angular momentum at the microscopic
level: Larmor gyration in the plasma case, particle spin in the chiral fluid case \citep{Lingam2015}. The
correspondence is structural rather than literal --- in the magnetised plasma the gyration is imposed by an
external field, whereas the spin considered here is generated by the activity itself and dissipated against a
substrate --- but in both cases a microscopic angular momentum forces a transverse, non-dissipative stress,
and the systematic derivation given below may therefore be of interest beyond the active-matter context.

\subsection{Structure of the work}
\label{sec:structure} 
The work is structured as follows. In \S\ref{sec:conservative_form} we write the conservative form of the
mass, linear-momentum and angular-momentum balance equations of the fluid, without any symmetry
assumption. In \S\ref{sec:convective_form} we deduce their convective form and show that two kinds of fluid
emerge: fluids of the kind I, which are subject to no net body couple and whose stress tensor is symmetric;
and fluids of the kind II, which, under the action of net body couples, develop additional contributions to
the stress tensor and are no longer fully symmetric. In \S\ref{sec:kind_I} we recall, for completeness, the
well-known form of the stress tensor for fluids with no net angular-momentum density. In \S\ref{sec:kind_II}
we classify the kind II fluids into two types according to the origin of the body couple acting on them, and
we deduce the complete form of the stress tensor and the spin flux for kind II fluids. In
\S\ref{sec:incompressible} we specialise the balance equations to incompressible flow at constant transport
coefficients. On this basis, \S\ref{sec:base_chiral} constructs the simplest steady base states of a
two-dimensional chiral fluid --- the quiescent states (\S\ref{sec:hydrostatic}) and the forced azimuthal
flows (\S\ref{sec:azimuthal}) --- and \S\ref{sec:cavity} treats the boundary-driven chiral Stokes cavity, in
both a circular domain (\S\ref{sec:circular}), where a closed-form solution is available, and a square domain
(\S\ref{sec:square}), solved numerically. A quantitative comparison of these predictions with experimental
data from a system of air-fluidised spinning disks \citep{Lopez_Castano_2022} is in preparation and is not
included in the present version. Finally, \S\ref{sec:conclusion} is devoted to the discussion of the results
and the conclusions.

\section{Conservative form of the balance equations for the chiral fluid}
\label{sec:conservative_form}

We consider the standard hydrodynamic fields for a Newtonian fluid: mass density ($\rho$) and flow velocity
($\boldsymbol{u}$). We will assume constant temperature ($T$) throughout the work. The particles within the
fluid volume element would be subject in general to the action of torques --- coming from particle and/or
boundary interactions --- which will make them rotate, producing an extra hydrodynamic field: the spin field
($\boldsymbol\Omega$). These particles may be molecules, for instance; or any other constituent particle unit
for the fluid. However, the average particle rotation in the fluid element, due to thermalization, will in
general be null and, as a consequence, the angular momentum field for a typical fluid is null everywhere. The
reason is that torques from particle--particle and particle--boundary interactions are typically stochastic
under both stationary and transient conditions in a thermalised fluid, and therefore randomly oriented in
space, yielding the reduction $\boldsymbol{\Omega} = \boldsymbol{0}$ valid for a large number of cases in
fluid dynamics \citep{Batchelor_1967}.
  
However, we are interested here in fluids whose particles are subject to torques that persistently act in a
given sense, so that particle angular momentum comes to predominate along a certain direction of space,
yielding $\boldsymbol{\Omega} \neq \boldsymbol{0}$. As noted above, a number of natural systems present this
property. Therefore, we take into consideration the balance equations for mass, linear
momentum \textit{and angular momentum} in a fluid --- the last of which is capable of sustaining a net
spin field. We write them here in their most fundamental conservative
form 

\begin{align}
  \label{eq:mass_conservative_form}
  & \partial_t \rho + \boldsymbol{\nabla} \bcdot (\rho\, \boldsymbol{u}) = 0\\
  \label{eq:linear_momentum_conservative_form}
  & \partial_t (\rho\, \boldsymbol{u}) + \boldsymbol{\nabla} \bcdot (\rho\, \boldsymbol{u}\otimes\boldsymbol{u})
    = \boldsymbol{f} + \boldsymbol{\nabla} \bcdot \boldsymbol\sigma\\
  \label{eq:angular_momentum_conservative_form}
  & \partial_t \!\left[\rho\, (\boldsymbol{r}\times\boldsymbol{u}) + \rho I\,\Omega \right] +
    \boldsymbol{\nabla} \bcdot \!\left\{ \!\left[\rho \,(\boldsymbol{r}\times\boldsymbol{u}) +
    \rho I\, \Omega\right] \boldsymbol{u} \right\}  =  \tau + \boldsymbol{\nabla}\bcdot \boldsymbol\Sigma,
\end{align}
where $\partial_t\equiv \partial/\partial t$ is the partial time derivative, $\otimes$ stands for tensor product
(for 2D vectors, $(\boldsymbol{u}\otimes\boldsymbol{u})_{ij} = u_iu_j$), and $\rho I$ represents moment of
inertia density, with $I$ being the micromoment of inertia, as defined by \citet{Eringen_1966}, and which we
consider to be constant in our fluid. Also, $\rho\boldsymbol{u}$ and $\rho I\, \boldsymbol\Omega$ denote
linear momentum and spin angular momentum density fields, respectively; in two dimensions the angular momentum
is an axial vector normal to the flow plane, and we write throughout its single (pseudoscalar) component
$\Omega$.\footnote{In writing the internal angular momentum as $\rho I\,\boldsymbol\Omega$ we assume that it
  reduces to particle spin alone. In general it may also receive a contribution from the motion of the
  particles about the hydrodynamic centre of the fluid element \citep{Klymko_2017}. That contribution is an
  independent field only when the particles sustain a circulation of their own --- as for swimmers on closed
  orbits of radius smaller than the coarse-graining scale --- and is otherwise nothing but the vorticity of
  the coarse-grained flow. The air-fluidised disks considered here are rigid bodies advected by the local
  flow, and no such contribution arises.}$^{,}$\footnote{For the same reason, the quantities
  $\boldsymbol{r}\times\boldsymbol{f}$, $\boldsymbol{r}\times\boldsymbol{u}$ and $\boldsymbol\tau$ likewise
  point along the $Z$ direction, and thus they can also be treated as pseudoscalars here.} In particular, we
consider the fluid to lie in the $XY$ plane, so that the axial vector $\boldsymbol\Omega$ points along the $Z$
direction. Note the torque
$\tau = \tau_f +\tau_0$ in the angular momentum balance, \eqref{eq:angular_momentum_conservative_form}, where
$\tau_f \equiv \boldsymbol{r} \times \boldsymbol{f}$ is the moment of the net body force $\boldsymbol{f}$ on
the fluid element, and $\tau_0$ encodes the body couple on the fluid element (see below). Finally,
$\boldsymbol\Sigma\equiv (\boldsymbol{r}\times\boldsymbol{\sigma}) + \boldsymbol{C}$ with $\boldsymbol{C}$
being the spin flux, denotes the surface contributions to angular momentum.

By body force $\boldsymbol{f}$ we mean here the total volumetric force acting on the fluid element,
including both externally applied fields (gravity, electromagnetic) and the friction $-\Gamma\boldsymbol{u}$
exerted by the substrate on the fluid. Correspondingly, the body couple $\tau_0$ is the total torque
density on the element that is not the moment of a force, and decomposes into three physically distinct
contributions,
\begin{equation}
  \label{eq:body_couple_decomp}
  \tau_0 \;=\; \tau_a \;+\; \tau_b \;+\; \tau_{\mathrm{drag}} , \qquad
  \tau_{\mathrm{drag}} \;\equiv\; -\,\Gamma^\Omega\,\Omega ,
\end{equation}
where $\tau_a$ is the intrinsic active torque generated by the particles' own microscopic dynamics,
$\tau_b$ is the torque exerted by an external non-scalar field coupled to particle orientation (e.g.\
$\tau_b = \boldsymbol{m}\times\boldsymbol{B}$ for a magnetic moment $\boldsymbol{m}$ in a field
$\boldsymbol{B}$), and $\tau_{\mathrm{drag}}$ is the rotational friction with the substrate. The drag
terms $-\Gamma\boldsymbol{u}$ and $-\Gamma^\Omega\Omega$ are the translational and rotational counterparts
of a single dissipative mechanism, internal to the composite fluid + substrate system rather than external
body actions in the classical sense. The classification of chiral active fluids developed in
\S\ref{sec:kind_II} rests on the origin of the active contributions $\tau_a, \tau_b$; the drag term
$\tau_{\mathrm{drag}}$ is universal.

In \eqref{eq:linear_momentum_conservative_form}--\eqref{eq:angular_momentum_conservative_form} body
($\boldsymbol{f}, \tau$) and surface contributions ($\boldsymbol\sigma, \boldsymbol\Sigma$) are of
qualitatively different origin. Body forces and torques are long ranged since they vary slowly in space. As a
consequence, they have a unique value within a fluid volume element; i.e., they yield well defined
force/torque fields within the entire fluid. On the contrary, surface forces are very short-range since they
decay rapidly with distance, in such a way that they become negligible at distances higher than the order of
magnitude of particle size. For this reason, they act only within a fluid layer; their origin lies in the
transport of momentum of the migrating particles that cross the interface of two fluid elements in direct
contact.\footnote{In the case of high particle density fluids, an additional contribution may arise from
  short-range interparticle forces near the interface of fluid elements in contact
  \citep{Batchelor_1967}.} These interactions are described by a set of transport coefficients whose values
depend on the physical properties of the fluid \citep{Chapman_1970,Kim_1998} and that appear in
$\boldsymbol{\nabla}\bcdot\boldsymbol\sigma$ and $\boldsymbol{\nabla}\bcdot\boldsymbol\Sigma$ when the fluxes
are expressed as functions of the average fields.

\section{Convective form of the balance equations for the chiral fluid}
\label{sec:convective_form}

From the conservative form of the balance equations
\eqref{eq:mass_conservative_form}--\eqref{eq:angular_momentum_conservative_form}, we can obtain the
corresponding convective form, which must be written carefully since we now include the angular momentum
balance as well.

\begin{align}
  & \frac{\mathrm{D}\rho}{\mathrm{D}t} + \rho \boldsymbol{\nabla\cdot u} = 0,
 \label{eq:mass_convective_form} \\
  \rho &\frac{\mathrm{D}\boldsymbol{u}}{\mathrm{D}t}  =  \boldsymbol\nabla\cdot\boldsymbol\sigma  + \boldsymbol{f},
 \label{eq:momentum_convective_form} \\
  \rho I &\frac{\mathrm{D}\Omega}{\mathrm{D}t} =  \boldsymbol{\nabla}\bcdot\boldsymbol{C} + 2 \sigma^* + \tau_0,
\label{eq:angular_momentum_convective_form}
\end{align} where $\boldsymbol{C}$ is the spin flux and
\begin{equation}
  \label{eq:sigma_star_def}
  \sigma^* \equiv \tfrac{1}{2}\, \mathrm{tr}\!\left(\epsilon\bcdot\boldsymbol\sigma\right)
           = \tfrac{1}{2}\,\epsilon_{ij}\,\sigma_{ji},
\end{equation} 
is the pseudoscalar dual of the stress tensor $\boldsymbol\sigma$ \citep[in fact, the Hodge dual of
$\boldsymbol\sigma$, see for instance][]{Frankel_2012}, which is in this case the two-dimensional analogue of
the axial vector of a skew tensor.\footnote{Here, $\sigma^*$ is mathematically a pseudoscalar field (a scalar
  with broken parity). An analogy: in 3D the vorticity vector would be the Hodge dual of the spin
  tensor.}  Here, $\epsilon$ is the 2D Levi-Civita tensor

\begin{equation}
  \label{eq:levi_civita_2D}
  \boldsymbol\epsilon \equiv \begin{bmatrix}
    0 & 1 \\
    -1 & 0
  \end{bmatrix}.
\end{equation}

Thus, the symmetric part of $\boldsymbol\sigma$ is blind to the Hodge contraction, since the operation
$\mathrm{tr}\!\left(\epsilon\bcdot\boldsymbol\sigma\right)$ extracts precisely the antisymmetric channel of
the stress; i.e. $\mathrm{Tr}(\boldsymbol\epsilon\cdot\boldsymbol\sigma)
=\mathrm{Tr}(\boldsymbol\epsilon\cdot\boldsymbol\sigma^{\mathrm{anti}})$.  
 
We will deduce the balance equations in convective form
\eqref{eq:mass_convective_form}--\eqref{eq:angular_momentum_convective_form} in
\S\ref{sec:mass_balance}--\ref{sec:angular_momentum_balance}, paying special attention to the angular momentum
balance. We will analyze in \S\ref{sec:kind_II} the types of fluids with non-vanishing $\tau_0$ according to
the nature of the torques present in the fluid.

\subsection{Mass balance}
\label{sec:mass_balance}

From \eqref{eq:mass_conservative_form} it is straightforward to obtain
  
\begin{equation}
  \label{eq:mass_balance}
\frac{\mathrm{D}\rho}{\mathrm{D}t} + \rho \boldsymbol{\nabla\cdot u} = 0,
\end{equation}
where $\mathrm{D}/\mathrm{D t} \equiv \partial /\partial t + \boldsymbol u \cdot \boldsymbol{\nabla}$
represents the material derivative, commonly used in fluid mechanics \citep{Batchelor_1967}. The material
derivative represents the time variation of a quantity as moving along within a fluid volume element that is
moving with velocity $\boldsymbol{u}$. So, an important class of fluid flows emerges when this
derivative is zero for the fluid mass density; the incompressible flows, for which the condition
$\boldsymbol{\nabla\cdot u} = 0$ follows from \eqref{eq:mass_balance} if $\mathrm{D}\rho/\mathrm{D}t=0$.

\subsection{Linear momentum balance}
\label{sec:momentum_balance}

Since
$\boldsymbol{\nabla}\bcdot(\rho\,\boldsymbol{u}\otimes\boldsymbol{u}) =
\boldsymbol{u}\boldsymbol{\nabla}\bcdot (\rho\,\boldsymbol{u}) +
(\rho\,\boldsymbol{u})\bcdot\boldsymbol\nabla\,\boldsymbol{u}$, we can write

\begin{equation}
  \label{eq:linear_momentum_conservative_force_reduced}
  \underbrace{\rho\,\partial_t\boldsymbol{u} +
    \rho\,\boldsymbol{u}\bcdot\boldsymbol{\nabla}\boldsymbol{u}}_{=\rho\frac{\mathrm{D}\boldsymbol{u}}{\mathrm{D}t}}
  + \underbrace{\boldsymbol{u} [\partial_t\rho +
    \boldsymbol{\nabla}\bcdot(\rho\,\boldsymbol{u})]}_{=0}  =
  \boldsymbol{\nabla}\bcdot{\boldsymbol\sigma} + \boldsymbol{f}. 
\end{equation}

Here we have used the mass conservation equation~\eqref{eq:mass_conservative_form}. Thus, we may obtain the
simplified and well-known convective form of the momentum balance \citep{Batchelor_1967}

\begin{equation}
  \label{eq:momentum_balance_convective_form}
  \rho\frac{\mathrm{D}\boldsymbol{u}}{\mathrm{D}t}  =  \boldsymbol\nabla\cdot\boldsymbol\sigma  + \boldsymbol{f}.
\end{equation}

The particular structure of the stress tensor is to be discussed in \S\ref{sec:odd_stress}, where we will
deduce a generalization of the stress tensor for the two-dimensional chiral fluid.

\subsection{Angular momentum balance. The symmetry properties of the stress tensor}
\label{sec:angular_momentum_balance}

Just as we used the conservative form of mass balance \eqref{eq:mass_conservative_form} into
\eqref{eq:linear_momentum_conservative_form}, we can iteratively use the latter into the conservative form of
angular momentum balance in order to simplify it. Moreover, we can obtain information about the symmetry
properties of the stress tensor out of this reduced version of angular momentum, as we will see.
 
For this purpose, note first that in~\eqref{eq:angular_momentum_conservative_form} we have written the angular
momentum of the fluid as the sum of two contributions. The first one ($\boldsymbol{r} \times \boldsymbol{u}$)
comes from the rotation of the fluid element as a whole. The second comes from the average of angular velocities
of the particles contained within the fluid volume element. We denote this average field as
$\Omega$, and call it the \textit{spin field}. It is worth stressing that both contributions
carry fluid angular momentum but their origin differs: the latter has its origin at particle level, whereas
the former is produced by the fluid volume element itself, as we said.

Typically, most fluids are composed of particles with only passive rotation, which arises from momentum
exchange upon collisional interactions. The stochastic nature of these rotations results in a spin field
that is zero everywhere in the fluid, even in the case of inherently non-equilibrium fluid states
\citep{Reyes_2014}. For this reason, the spin field vanishes by default in most fluids. As is well known, a
straightforward consequence of this situation is a completely symmetric stress tensor. This follows from the
result of zero angular momentum balance in the rotationally passive fluid \citep{Batchelor_1967}.

However, if we need to analyze a more generic fluid, including for instance biological systems as discussed in
\S\ref{sec:intro}, we need to consider the possibility $\Omega \neq 0$, and then a
more complex structure of $\boldsymbol\sigma$ follows, which now is not completely symmetric (nor antisymmetric) as we
demonstrate in this section. In effect, the total 2D fluid angular momentum may be represented as

\begin{equation}
  \label{eq:L_definition}
  L = \int \mathrm{d}V \rho\left[(\boldsymbol{r}\times \boldsymbol{u}) + I \Omega \right] = \int
  \mathrm{d} V \rho\left(\epsilon_{ij} \,r_iu_j + I\Omega\right), 
\end{equation}
and its total time derivative must be, according to Newton's second law, equal to the sum of the couples
acting on the fluid. These decompose into surface contributions, collected in the angular momentum flux
$\boldsymbol{\Sigma} = \boldsymbol{r} \times \boldsymbol\sigma + \boldsymbol{C}$ of
\eqref{eq:angular_momentum_conservative_form}, and volume contributions
($\boldsymbol{r} \times \boldsymbol{f} + \tau_0$). The second surface contribution ($\boldsymbol{C}$)
transports the angular momentum associated with particle rotation within the fluid volume element, whereas the
first ($\boldsymbol{r} \times \boldsymbol\sigma$) is the moment of the tractions and transports the angular
momentum carried by the motion of the fluid element itself. Thus
  
\begin{align}
  \label{eq:L_time_derivative}
    \frac{\mathrm{d}L}{\mathrm{d}t} &=  \int\mathrm{d} \boldsymbol{S} \bcdot \,\boldsymbol{\Sigma} + \int
    \mathrm{d}V\, (\boldsymbol{r}\times \boldsymbol{f} + \tau_0) = \int\mathrm{d}
    \boldsymbol{S}\bcdot (\boldsymbol{r}\times\boldsymbol{\sigma}) + \int\mathrm{d} \boldsymbol{S} \bcdot
    \,\boldsymbol{C}  \\
                                    & + \int \mathrm{d}V\, (\boldsymbol{r}\times \boldsymbol{f} +\tau_0 ) =
    \int\mathrm{d}S\,(\epsilon_{ij}r_i\sigma_{jl}n_l + C_{l}n_l) +\int\mathrm{d}V (\epsilon_{ij}r_i f_j + \tau_0)
   . \nonumber
\end{align}
The surface integrals in the above equation can be further manipulated, applying Gauss theorem, to
yield
\begin{equation}
    \int  \mathrm{d} V \partial_l(\epsilon_{ij}r_i\sigma_{jl}) + \int\mathrm{d} V \partial_lC_{l}
    = \int\mathrm{d} V \epsilon_{ij}r_i\partial_l\sigma_{jl} + \int\mathrm{d}V \epsilon_{ij}\sigma_{ji} +
     \int\mathrm{d} V \partial_lC_{l}, \label{eq:L_right_hand}
\end{equation}
where we have taken into account that $\partial_lr_i = \delta_{li} = \delta_{il}$ and thus
$(\partial_lr_i)\sigma_{jl} = \delta_{li}\sigma_{jl} = \sigma_{ji}$. Thus, we may rewrite
\eqref{eq:L_time_derivative} as

\begin{equation}
  \label{eq:L_time_derivative_2}
 \frac{\mathrm{d}L}{\mathrm{d}t} = \int\mathrm{d} V \epsilon_{ij}r_i\partial_l\sigma_{jl} + \int\mathrm{d}V \epsilon_{ij}\sigma_{ji} +
     \int\mathrm{d} V \partial_lC_{l} + \int\mathrm{d}V (\epsilon_{ij}r_i f_j + \tau_0).
 \end{equation}
Also, taking the time derivative of \eqref{eq:L_definition} we may write 
\begin{align}
  \nonumber
  & \frac{\mathrm{d}L}{\mathrm{d}t } = \int{\mathrm{d}V }
    \left[\partial_t(r_i\epsilon_{ij}\rho\, u_j) + \partial_k(r_i\epsilon_{ij}\rho\, u_ju_k)\right]  + \int
    \mathrm{d}V \left[\partial_t(\rho I\Omega ) + \partial_k(\rho I\Omega u_k) \right] \\
&   =  \int{\mathrm{d}V }  \left[r_i\epsilon_{ij}\partial_t(\rho\, u_j) +  r_i\epsilon_{ij} \partial_k(\rho\, u_ju_k)\right]
    + \int \mathrm{d}V I\Omega [\underbrace{\partial_t\rho + \partial_k(\rho u_k)}_{=0}] \nonumber \\
  &+ \int\mathrm{d}V \rho [\underbrace{\partial_t(I\Omega) + u_k \partial_k(I\Omega)}_{=\frac{\mathrm{D}(I\Omega)} {\mathrm{D} t}}
]  \label{eq:L_conservation_3}
\end{align}
since $\partial_t + u_k\partial_k \equiv \frac{\mathrm{D }}{\mathrm{D}t}$ and we took into account that
$\partial_k(r_i\epsilon_{ij})(\rho u_ju_k) = \epsilon_{ij}\delta_{ki}(\rho u_ju_k) = \rho\epsilon_{ij}u_ju_i
=0$ since the full contraction of a totally antisymmetric tensor ($\epsilon_{ij}$) with a symmetric tensor
($u_iu_j$) is always zero. Also, $(\partial_t r_i)\,\epsilon_{ij}\,\rho u_j = 0$ identically, since the
Eulerian coordinates $r_i$ are independent of time. Thus, combining the right side of the equality in
\eqref{eq:L_right_hand} and the last equality in \eqref{eq:L_conservation_3} we may write,
 
\begin{align}
 \label{eq:L_conservation_4}
  \int{\mathrm{d}V }
 r_i\epsilon_{ij} & \underbrace{\left[\partial_t(\rho\, u_j) + \partial_k(\rho\, u_ju_k) -  \partial_l\sigma_{jl}- f_j\right]}_{=0} \\
  &  + \int \mathrm{d}V
    \left[\rho\frac{\mathrm{D}(I\Omega)}{\mathrm{D}t}-\boldsymbol{\nabla}\bcdot\boldsymbol{C} - \tau_0\right] =  \int\mathrm{d}V \epsilon_{ij}\sigma_{ji}. \nonumber
\end{align}
Notice also that the bracketed term in \eqref{eq:L_conservation_4} vanishes due to linear momentum
balance~\eqref{eq:linear_momentum_conservative_form}. Thus, Equation~\eqref{eq:L_conservation_4} is much
reduced, yielding
 
\begin{equation}
  \label{eq:L_conservation_5}
  \int \mathrm{d}V \left[ \rho\frac{\mathrm{D}(I\Omega)}{\mathrm{D}t}
    -\boldsymbol{\nabla}\bcdot\boldsymbol{C} - \tau_0 \right]=  \int\mathrm{d}V \epsilon_{ij}\sigma_{ji}.
\end{equation}

Equation \eqref{eq:L_conservation_5} is an important result, since it has two fundamental and mutually
exclusive implications

\begin{enumerate}[label=\Roman*.]
\item If the fluid has no sources of angular momentum field, i.e.\ if $\tau_0=0$, then
  $\Omega = 0$ and $\boldsymbol{C}=\boldsymbol{0}$, and \eqref{eq:L_conservation_5} yields
 
\begin{equation}
  \label{eq:P_is_symmetric}
 \int\mathrm{d}V \epsilon_{ij}\sigma_{ji} = 0 \quad \Rightarrow \quad \sigma_{ij} = \sigma_{ji}
\end{equation}
\item If otherwise the fluid has a non-null body couple $\tau_0\neq 0$, then
  \begin{equation} 
  \label{eq:P_is_not_symmetric}
 \int\mathrm{d}V \epsilon_{ij}\sigma_{ji} \neq 0 \quad \Rightarrow \quad \sigma_{ij} \neq \sigma_{ji}.
\end{equation}
\end{enumerate}  

Note that $\sigma_{ij} \neq \sigma_{ji}$ only implies that the stress tensor is not fully
symmetric. This does not imply that $\boldsymbol\sigma$ is antisymmetric either. Therefore, we might write a
generalized stress tensor as the sum of a symmetric and an anti-symmetric part:
$\boldsymbol\sigma = \boldsymbol\sigma^{\mathrm{sym}} + \boldsymbol\sigma^{\mathrm{anti}}$. It is actually not
guaranteed that $\boldsymbol\sigma^\mathrm{sym}$ equals the traditional full form of the stress tensor for
Newtonian fluids with no spin field. However, taking into account the property
$\epsilon_{ij}\sigma^{\mathrm{sym}}_{ij} = 0$ in \eqref{eq:L_conservation_5}, and noting that the
integration volume is arbitrary, we may readily write the local balance

\begin{equation}
  \label{eq:L_balance_chiral}
  \rho\frac{\mathrm{D}( I\Omega)}{\mathrm{D}t} - \boldsymbol{\nabla}\bcdot\boldsymbol{C} - \tau_0=
  \epsilon_{ij} \sigma_{ji} = 2\sigma^*,
\end{equation}
where we made use of the Hodge dual $\sigma^*$, as defined in equation
\eqref{eq:sigma_star_def}.

A caveat is in order at this point. \citet{Markovich_2019} emphasise that only the part of the antisymmetric
stress which cannot be written as a divergence acts as a body couple in the spin balance: a contribution of
divergence form transmits a surface torque instead, and inserting it in \eqref{eq:L_balance_chiral} yields an
incorrect description. In two dimensions the antisymmetric channel is spanned by $\epsilon_{ij}$ alone, so
the question reduces to whether its pseudoscalar coefficient is itself a divergence. It is not: as shown in
\S\ref{sec:chiral_pressure_form}, that coefficient is the chiral pressure $p_R = \mu_R(\Omega-\omega/2)$,
built from the spin field $\Omega$, which obeys its own balance equation and is not slaved to any vector
potential. The whole of $\boldsymbol\sigma^{\mathrm{anti}}$ therefore enters \eqref{eq:L_balance_chiral}
legitimately as a body couple.

\section{Fluids of the kind I: the symmetric stress tensor}
\label{sec:kind_I}

The standard Newtonian stress is usually expressed as the sum of two contributions. The first is an isotropic
tension, equal along all directions of the fluid volume element; at rest it reduces to the hydrostatic
pressure $p$, which defines the hydrostatic (flowless) states of the fluid. The second contribution expresses
the deviations from the isotropic one and contains off-diagonal elements; for this reason, it has full
rank-two tensor form. It is typically called the \textit{deviatoric} stress and it vanishes for fluids at
rest. Its physical origin lies in the friction forces between fluid regions in relative motion across the
interface of a volume element. These friction forces are characterised by the spatial variations of the flow
velocity; for this reason, in a first approximation, the deviatoric stress is expressed as a linear and
symmetric combination of the first spatial derivatives of the flow velocity $\boldsymbol{u}$. Thus, we write  
  
\begin{align} 
  &   \sigma_{ij}^\mathrm{I}= -p\,\delta_{ij} + \zeta\,(\boldsymbol{\nabla}\bcdot\boldsymbol{u})\,\delta_{ij} +
    2\mu\, d_{ij}\, , \label{eq:Newton_stress}\\
  & \text{with} \quad d_{ij} \equiv \left( e_{ij} - \frac{1}{2}(\boldsymbol{\nabla\cdot u})\, \delta_{ij}\right)\quad
    \text{and} \quad  e_{ij} \equiv \frac{1}{2}\left(\frac{\partial u_i}{\partial r_j} +
    \frac{\partial u_j}{\partial r_i}\right), \label{eq:rate_of_strain}
\end{align}
where $d_{ij}$ is the \textit{deviatoric strain-rate tensor}, $p$ stands for the hydrostatic pressure and
$\zeta \geq 0$ is the bulk (expansion) viscosity. This is the classical decomposition of the stress tensor
in fluid mechanics \citep{Batchelor_1967}, splitting the stress into two mechanical channels: the isotropic
channel, which carries both the equilibrium pressure and a viscous resistance to compression/dilation; and
the \textit{deviatoric} channel $d_{ij}$, which emerges solely from the shearing of the velocity field
$\boldsymbol{u}$. The deviatoric stress is traceless and symmetric; its contributions tend to deform the
fluid volume element, and this deformation is realised through fluid motion, hence inducing flow. The bulk
term vanishes identically for incompressible flow. We have also defined the rate-of-strain tensor $e_{ij}$
in \eqref{eq:rate_of_strain}.
 
It is mathematically straightforward to check that the complete stress tensor $\boldsymbol\sigma^\mathrm{I}$ is indeed
completely symmetric --- it is a linear combination of the symmetric tensors $\delta_{ij}$ and $d_{ij}$, so
that $\sigma_{ij}^\mathrm{I} = \sigma_{ji}^\mathrm{I}$ --- in agreement with the result of
\S\ref{sec:angular_momentum_balance} for fluids with no body couple ($\tau_0 = 0$).

\section{Fluids of the kind II: the odd stress tensor, the spin flux, and the body couple}
\label{sec:kind_II}

According to the derivation of \S\ref{sec:angular_momentum_balance}, what distinguishes the kind-II fluid from
the kind-I fluid is a single primitive input of the theory: the body couple $\tau_0$, which is present in the
angular-momentum balance \eqref{eq:angular_momentum_conservative_form}. The continuum framework accommodating
such couples is classical \citep{Dahler_1961,Condiff_1964,Eringen_1966}; what varies across physical systems
is the mechanism that produces them. We recall that a polar body force such as gravity cannot play this role:
it enters the angular-momentum budget only through its moment $\boldsymbol{r}\times\boldsymbol{f}$, which
cancels identically against the linear-momentum balance in the reduction of
\S\ref{sec:angular_momentum_balance}, and therefore never sources the spin field. A body couple therefore
requires a coupling to the internal structure of the particles to emerge in the angular-momentum balance
\eqref{eq:angular_momentum_convective_form}. Two physically distinct mechanisms provide such a coupling, each
defining a subclass of kind-II fluid.

Following the decomposition \eqref{eq:body_couple_decomp} introduced in \S\ref{sec:conservative_form},
the rotational drag $\tau_{\mathrm{drag}}$ is universal --- present in every chiral fluid coupled to a bath
--- whereas it is the origin of the active contributions $\tau_a$ and $\tau_b$ that distinguishes the two
subclasses of kind-II fluid.

Both $\tau_a$ and $\tau_b$ are genuine body couples --- torques exerted on the internal rotational degree
of freedom, with vanishing net force --- and therefore neither is expressible as the moment
$\boldsymbol{r}\times\boldsymbol{f}$ of a resultant body force. A uniform field makes the point sharp: it
exerts $\tau_b = \boldsymbol{m}\times\boldsymbol{B}\neq\boldsymbol{0}$ while exerting no net force at
all. The distinction between them is not mechanical but one of origin --- intrinsic activity versus
external-field coupling --- and it is precisely this that defines the two subclasses developed below.

In the first subclass, which we denote \textit{kind II-a} --- the one addressed in this work --- the couple
originates in the internal dynamics of the particles themselves: chemo-mechanical motors in the biological
systems of \S\ref{sec:intro}, the mechanical rectification of vertical vibration by chirally shaped grains
\citep{Tsai_2005} , or the persistent spin in bladed-disks in the experiments used for comparison below
\citep{Lopez_Castano_2022}, or actively torqued grains in particle simulations \citep{Han_2021}. No
macroscopic field participates: $\tau_b = 0$, so that $\tau_0 = \tau_a + \tau_{\mathrm{drag}}$ with the active
couple $\tau_a$ a prescribed material source; the handedness is a property of the particle rather than of the
environment, and the theory closes on the hydrodynamic fields alone, as in \S\ref{sec:spin_flux}.
 
In the second subclass, \textit{kind II-b}, the couple is imposed from outside by a field of axial
character. The paradigm is the magnetic suspension driven by a rotating field, for which
$\tau_b = \boldsymbol{M}\times\boldsymbol{B}$ (a pseudoscalar in two dimensions), with $\boldsymbol{M}$ the
magnetisation density; magnetised plasmas, where gyroviscosity plays the role of the odd stress
\citep{Braginskii1965}, and particle suspensions undergoing electrorotation, described within the
leaky-dielectric framework \citep{Melcher_1969,Saville_1997}, belong to the same subclass. Here $\tau_0$ is
not a datum but a function of state: the magnetisation obeys its own relaxation equation, advected and rotated
by the flow \citep{Shliomis_1972}, the internal field must be determined self-consistently from the
magnetostatic problem --- non-local and geometry-dependent --- and the field enters the linear momentum
balance as well, through the Kelvin force and the Maxwell stress \citep{Rosensweig_1985}. The hydrodynamics of
a kind II-b fluid is therefore bidirectionally coupled to a field problem, in the manner of electro
\citep{VegaReyes_2002}- and magnetohydrodynamics \citep{Chandrasekhar_1981}. The canonical base state of the
subclass --- the spin-up flow of a ferrofluid in a uniform rotating field \citep{Moskowitz_1967} ---
illustrates the price of this coupling: its explanation remained in dispute for over five decades, and the
dispute concerned precisely the boundary conditions on the spin field and the dynamical role of the
couple-stress term $\kappa\nabla^2\Omega$ \citep{Zaitsev_1969,Rosensweig_1990,Chaves_2006,Shliomis_2021} ---
the two ingredients that, for the fluids considered in this work, are settled by the microscopic estimate
$\ell_s \lesssim a$ of \S\ref{sec:spin_flux}. A further signature of the subclass is that the handedness is
inherited from the field: reversing the field reverses the chiral response, so that the odd coefficients are
odd functions of the drive, in the sense of Onsager--Casimir reciprocity.

The distinction between the subclasses is not always sharp, and usefully so: when the external rotating field
is imposed \textit{hard} --- weak susceptibility, negligible feedback of the flow on the field --- the kind
II-b equations collapse onto the kind II-a structure, with $\tau_0$ effectively prescribed. Most tabletop
realisations of chiral liquids made of magnetically driven spinners operate in this regime
\citep{Soni_2019}. In this sense, the equations derived here for the intrinsically active subclass constitute
the reduced description of confined chiral flow, upon which kind II-b superposes field dynamics only when the
feedback is significant.

\subsection{The odd stress tensor}
\label{sec:odd_stress}

For fluids of the kind II, as we have shown, the existence of a non-vanishing spin field precludes in
general a symmetric stress tensor; the stress has no definite symmetry property, and must be decomposed
into symmetric and antisymmetric parts.

It is reasonable to expect that kind-II fluids remain subject to the mechanical actions embodied by the kind-I
stress tensor $\boldsymbol\sigma^{\mathrm{I}}$: they will in general also undergo compression/dilation and
deformation/dissipation.  But, in addition, kind-II fluids experience a new set of mechanical actions stemming
from \textit{rotated} stresses, since their origin lies in the body couple $\tau_0$. We will represent
rotations here by means of the 2D Levi-Civita tensor $\epsilon_{ij}$, as written in equation
\eqref{eq:levi_civita_2D}.

Therefore, the stress tensor of the kind-II fluid is the sum of $\boldsymbol\sigma^{\mathrm{I}}$ and a new contribution,
coined the \textit{odd stress}, which we write as the rotated version of $\boldsymbol\sigma^{\mathrm{I}}$:

\begin{equation}
  \label{eq:sigmaII}
  \boldsymbol\sigma^{\mathrm{II}} = \boldsymbol\sigma^{\mathrm{I}} + \boldsymbol\sigma^{\mathrm{odd}},
\end{equation}
where
\begin{equation}
  \label{eq:odd_stress}
  \sigma^{\mathrm{odd}}_{ij}
  = \epsilon_{ik}\left[\, -\sigma^*\,\delta_{kj} + 2\mu^{\mathrm{odd}}\, d_{kj}\,\right]
  = p_R\,\epsilon_{ij} + \zeta^{\mathrm{odd}}(\boldsymbol\nabla\cdot\boldsymbol{u}) \epsilon_{ij}+
  2\mu^{\mathrm{odd}}\,\epsilon_{ik}\, d_{kj} , 
\end{equation}
and we recall that $d_{ij}$ is the deviatoric strain-rate tensor of \eqref{eq:rate_of_strain}. In Equation
\eqref{eq:odd_stress} we have used
$-\sigma^* = p_R + \zeta^{\mathrm{odd}}(\boldsymbol\nabla\cdot\boldsymbol{u})$ since according to Equation
\eqref{eq:L_balance_chiral} $\sigma^*$ extracts the antisymmetric part of $\boldsymbol\sigma$, which should
come from the isotropic component \eqref{eq:Newton_stress}, since $\epsilon_{ik}d_{kj}$ is trivially
symmetric.  The first form in Equation \eqref{eq:odd_stress} makes the construction explicit: the bracket has
exactly the structure of $\boldsymbol\sigma^{\mathrm{I}}$ in \eqref{eq:Newton_stress}, with
$-p + \zeta(\boldsymbol\nabla\cdot\boldsymbol{u})\to -\sigma^*$ and $\mu \to \mu^{\mathrm{odd}}$, and the
rotation $\epsilon_{ik}$ acts on it as a whole. The construction mirrors the particle level: in microscopic
models of chiral active matter, chirality enters through transverse pairwise forces --- the central
interaction rotated by $90^{\circ}$ \citep{Caprini_2025} --- and the odd stress implements the same rotation
one level up, on the stress tensor itself.  The second, expanded form --- written, as
$\boldsymbol\sigma^{\mathrm{I}}$, in terms of the deviatoric tensor $d_{ij}$ of \eqref{eq:rate_of_strain} ---
is the one we use throughout. The bulk channel of $\boldsymbol\sigma^{\mathrm{I}}$ rotates in the same way,
$\zeta \to \zeta^{\mathrm{odd}}$, producing the antisymmetric contribution
$\zeta^{\mathrm{odd}}(\boldsymbol{\nabla}\bcdot\boldsymbol{u})\,\epsilon_{ij}$. Note also the consistency of
the construction: taking the pseudoscalar dual \eqref{eq:sigma_star_def} of the full tensor
\eqref{eq:sigmaII}, $\tfrac{1}{2}\epsilon_{ij}\sigma^{II}_{ji}$, the $\boldsymbol\sigma^{\mathrm{I}}$ part
vanishes by symmetry, and the $\mu^{\mathrm{odd}}$ term vanishes as well because $\epsilon_{ik}d_{kj}$ is
symmetric for traceless symmetric $d_{ij}$; the only surviving term is
$\tfrac{1}{2}\,\epsilon_{ij}\,\sigma^{\mathrm{II}}_{ji} = -p_R -
\zeta^{\mathrm{odd}}(\boldsymbol\nabla\cdot\boldsymbol{u}) = \sigma^*$. Finally, it is worth to mention that
the rotational pressure $p_R$ is a pseudoscalar field, since it was defined out of the pseudoscalar dual
$\sigma^*$\footnote{Under a proper rotation, $\sigma^*$ behaves like a normal scalar. However, under a spatial
  reflection (an improper rotation, such as flipping one axis), the Levi-Civita symbol $\epsilon_{ij}$ changes
  sign because it is an isotropic tensor only under proper rotations. Since $\sigma_{ij}$ is a physical tensor
  (its components transform properly under reflections), the coefficient $\sigma^*$ must flip its sign under
  reflection to compensate for $\epsilon_{ij}$. This is the exact definition of a pseudoscalar (or axial
  scalar). In fact, in 2D fluids, this is exactly how antisymmetric stress components or the 2D vorticity
  tensor ($\omega_{ij} = \tfrac{1}{2}\epsilon_{ij}\omega$) are mapped to a scalar rotation field.}.

It is instructive to write the odd stress as a matrix,

\begin{equation}
  \label{eq:odd_stress_matrix}
  \boldsymbol\sigma^{\mathrm{odd}} \equiv \begin{bmatrix}
    2\mu^{\mathrm{odd}}d_{xy} & p_R + \zeta^{\mathrm{odd}}(\boldsymbol\nabla\cdot\boldsymbol{u}) + 2 \mu^{\mathrm{odd}} d_{yy} \\
    -p_R -\zeta^{\mathrm{odd}}(\boldsymbol\nabla\cdot\boldsymbol{u}) - 2 \mu^{\mathrm{odd}} d_{xx} & -2\mu^{\mathrm{odd}}d_{xy}
  \end{bmatrix}.
\end{equation}

Also, the distinction between $p$ and $p_R$, and between $\mu$ and $\mu^{\mathrm{odd}}$, is justified since
they prescribe distinct mechanical actions. This is best illustrated by an example. Specifically, the odd
viscosity $\mu^\mathrm{odd}$ inherits contributions also from perpendicular fluid forces. Let us consider the
$XY$ component in the symmetric stress tensor $\boldsymbol\sigma^{\mathrm{I}}$ \eqref{eq:Newton_stress}. This
stands for the traction along the $X$ direction acting on a surface perpendicular to the $Y$ direction, and it
comes from a strain rate of the type $\partial_yu_x$ (thus, a shear traction is produced by a shear strain
rate). In the odd stress, however, the $XY$ component comes from a compressional/expansional strain rate of
the form $d_{yy}$ plus the rotational pressure and the bulk viscosity term,
$\sigma_{xy}^\mathrm{odd} = p_R + \zeta^{\mathrm{odd}}(\boldsymbol\nabla\cdot\boldsymbol{u}) +
2\mu^{\mathrm{odd}}d_{yy}$; i.e., a \textit{pressure}-type strain rate ($d_{yy}$) induces shear through the
action of $\mu^\mathrm{odd}$. Additionally, if the deviatoric component $d_{yy}$ and
$\boldsymbol\nabla\cdot\boldsymbol{u}$ are null, then $\sigma^{\mathrm{odd}}_{xy}$ still survives due to the
presence of $p_R$; i.e., $p_R$ induces shear. A very special kind of shear though, since it is 
\textit{deformation-free} (because $p_R$ is inherited from the isotropic part of the stress tensor).

\subsubsection{The odd force as a rotated force: why $\mu^\mathrm{odd}$ drives no circulation}
\label{sec:odd_rotated_force}
 
The previous example admits a companion that completes the picture. There, a compressional
strain rate produced a tangential traction.  Conversely, the shear $d_{xy}$ --- which in the symmetric sector
produces the familiar tangential drag $\sigma_{xy} = 2\mu\, d_{xy}$ --- produces in the odd sector a
\textit{normal} traction, since $ \sigma^\mathrm{odd}_{xx} = 2\mu^\mathrm{odd} d_{xy}$.

A pure shear generates a difference of normal stresses, reactive rather than viscoelastic in origin. Together,
both examples show that the odd stress answers every mechanical question a quarter-turn off: whatever traction
the Newtonian response would exert, the chiral fluid adds the same traction rotated by $90^\circ$ --- the
continuum version of the Lorentz-like microscopic forces invoked above, which deflect at right angles to what
provokes them.

One might be tempted to conclude that, since the odd response returns along the very direction of the momentum
exchange, it simply ``cannot twist the fluid'', and that this is why $\mu^\mathrm{odd}$ is absent from the
vorticity balance. The conclusion is correct but the reasoning is not: anisotropic normal stresses do, in
general, exert torques on fluid elements.\footnote{A stress field $\sigma_{xx}=-\sigma_{yy}=s(x,y)$, with no
  shear components at all, produces the force $\boldsymbol{f}=(\partial_x s,\,-\partial_y s)$, whose curl,
  $-2\,\partial_x\partial_y s$, does not vanish in general. Being ``normal'' does not mean being
  ``irrotational''; only the \textit{isotropic} normal stress $-p\delta_{ij}$ is torque-free by construction.}
The true mechanism is prettier, and it is special to two dimensions: rotating every vector of a field by
$90^\circ$ exchanges its two differential characters,
\begin{equation}
  \label{eq:div_curl_swap}
  \boldsymbol{\nabla}\times(\epsilon\bcdot\boldsymbol{a})
     = -\,\boldsymbol{\nabla}\bcdot\boldsymbol{a},
  \qquad
  \boldsymbol{\nabla}\bcdot(\epsilon\bcdot\boldsymbol{a})
     = \boldsymbol{\nabla}\times\boldsymbol{a},
\end{equation}
both immediate from $\epsilon_{mi}\epsilon_{ik}=-\delta_{mk}$. A quarter-turn converts torque content into
compression content, and vice versa. Now, the odd force is precisely a rotated viscous force: because the
deviatoric strain rate satisfies $\partial_j d_{kj} = \tfrac{1}{2}\nabla^2 u_k$ identically (compressible or
not), the divergence of the deviatoric odd stress is
\begin{equation}
  \label{eq:odd_force_rotated}
  f^\mathrm{odd}_i
   = 2\mu^\mathrm{odd}\,\partial_j\!\left(\epsilon_{ik} d_{kj}\right)
   = \mu^\mathrm{odd}\,\epsilon_{ik}\nabla^2 u_k
   = \frac{\mu^\mathrm{odd}}{\mu}\,
     \left(\epsilon\bcdot\boldsymbol{f}^{\,\mathrm{visc}}\right)_i ,
\end{equation}
the ordinary viscous force turned by $90^\circ$ and relabelled. By \eqref{eq:div_curl_swap}, its curl is minus
the divergence of the viscous force,
\begin{equation}
  \label{eq:odd_curl}
  \boldsymbol{\nabla}\times\boldsymbol{f}^{\,\mathrm{odd}}
   = -\,\mu^\mathrm{odd}\,\boldsymbol{\nabla}\bcdot(\nabla^2\boldsymbol{u})
   = -\,\mu^\mathrm{odd}\,\nabla^2(\boldsymbol{\nabla}\bcdot\boldsymbol{u}).
\end{equation}
All the twisting the odd stress could deliver is exactly the compressing that the ordinary viscous stress
cannot: incompressibility, not alignment, is what silences it. Equation \eqref{eq:odd_curl} also states when
it reawakens --- the moment the flow compresses, $\mu^\mathrm{odd}$ re-enters the vorticity balance as a
genuine source, $-\mu^\mathrm{odd}\nabla^2(\boldsymbol{\nabla} \bcdot\boldsymbol{u})$.

Where does the force go instead? In two-dimensional incompressible flow
$\nabla^2\boldsymbol{u} = -\,\epsilon\bcdot\boldsymbol{\nabla}\omega$, so that the viscous and odd forces form
a dual pair,
\begin{equation}
  \label{eq:dual_pair}
  \boldsymbol{f}^{\,\mathrm{visc}} = -\,\mu\,\epsilon\bcdot\boldsymbol{\nabla}\omega,
  \qquad
  \boldsymbol{f}^{\,\mathrm{odd}}  = \mu^\mathrm{odd}\,\boldsymbol{\nabla}\omega :
\end{equation}
the viscous force pushes \textit{along} the vorticity contours, the odd force pushes \textit{across} them ---
the same field, a quarter-turn apart. Being a pure gradient, the odd force is absorbed into the pressure
without residue,
\begin{equation}
  \label{eq:pressure_shift}
  p \;\longrightarrow\; p - \mu^\mathrm{odd}\,\omega .
\end{equation}
Odd viscosity is therefore not absent from the confined chiral fluid; it is hidden in the pressure field ---
alive in the normal tractions it exerts on the walls, and in the weakly compressible signatures it induces,
such as the density excess in vortex cores \citep{Banerjee2017} --- while leaving no fingerprint on $\psi$ or
$\omega$. Any reconstruction of the pressure from measured flow fields must use the shifted pressure
\eqref{eq:pressure_shift}.

The reactive character suggested by the Lorentz analogy is in fact exact. In the two-dimensional space of
shears --- spanned by the two channels $d_{xx}=-d_{yy}$ and $d_{xy}$ --- the map $d \mapsto \epsilon\bcdot d$
is again a rotation by $90^\circ$, so the odd stress is always orthogonal to the strain rate that produces it
and performs no work:
\begin{equation}
  \label{eq:no_work}
  2\mu^\mathrm{odd}\left(\epsilon_{ik} d_{kj}\right) e_{ij} \equiv 0 ,
\end{equation}
an identity for traceless symmetric $d$. Like the magnetic force on a moving charge, the odd stress deflects
and never brakes: $\mu^\mathrm{odd}$ is a non-dissipative, gyroscopic transport coefficient, and its sign is
not constrained by the second law.

Finally, the identity \eqref{eq:div_curl_swap}, used once more, is also the reason for the silence of
$\kappa^\mathrm{odd}$ that we will describe in \S\ref{sec:spin_flux}: the chiral spin flux is the rotated gradient
$\kappa^\mathrm{odd}\,\epsilon\bcdot\boldsymbol{\nabla}\Omega$, and the divergence of a rotated field is the
curl of the original --- here the curl of a gradient, which vanishes identically. The structural parallelism
between $\{\mu,\mu^\mathrm{odd}\}$ and $\{\kappa,\kappa^\mathrm{odd}\}$ is thus a single two-dimensional
operator identity applied twice, with one instructive asymmetry: the odd spin flux is silent unconditionally,
whereas the odd viscous force is silent conditionally on incompressibility.

\subsection{The constitutive form of the chiral pressure}
\label{sec:chiral_pressure_form}

The construction of \S\ref{sec:odd_stress} identifies $p_R$ (together with the odd compression/dilation term,
that will be absent in incompressible flow) as the pseudoscalar dual of the stress, but leaves
its dependence on the hydrodynamic fields unspecified. That dependence is fixed, as for every term of a
Newtonian constitutive relation, by the requirement that $p_R$ be a pseudoscalar, linear in the fields, that
vanishes in the appropriate reference state; and the parallel with the ordinary pressure fixes it
uniquely. The isotropic part of $\boldsymbol\sigma^{\mathrm{I}}$, $-p\,\delta_{ij}$, is conjugate to the isotropic
(volumetric) part of the rate of strain, whose measure is the trace
$e_{kk} = \boldsymbol{\nabla}\bcdot\boldsymbol{u}$: the pressure is the scalar response to local
dilatation. Under the rotation $\delta_{ij}\mapsto\epsilon_{ij}$ that generates the odd stress in
\S\ref{sec:odd_stress}, the isotropic channel maps onto the antisymmetric one, so that $p_R$ must be the
pseudoscalar response to the \emph{rotational} counterpart of dilatation. Two independent rotation rates are
available in the chiral fluid. The antisymmetric part of the velocity gradient is
$\omega_{ij} = \tfrac{1}{2}\epsilon_{ij}\omega$, whose single invariant is the vorticity $\omega$; accordingly
the local angular velocity of a fluid element is $\tfrac{1}{2}\omega$, not $\omega$
\citep{Batchelor_1967}. The spin field $\Omega$ supplies the second rate, that of the particles themselves.
The pseudoscalar measuring the rotational mismatch is their difference, $\Omega - \tfrac{1}{2}\omega$ --- the
relative angular velocity between the embedded particles and the fluid element that carries them.

Its distinguished status follows from a single physical requirement. In a rigid-body co-rotation of the whole
system the particles spin at exactly the rate of the surrounding fluid element, $\Omega = \tfrac{1}{2}\omega$,
so that no relative rotation exists and no chirality-driven exchange of angular momentum can occur; the chiral
pressure must then vanish. Of the linear pseudoscalars that can be formed from $\Omega$ and $\omega$, only the
combination $\Omega - \tfrac{1}{2}\omega$ possesses this property --- $\Omega$ alone, or $\Omega - \omega$,
would drive a spurious odd stress in a state of uniform rigid rotation. This selects
\begin{equation}
  \label{eq:chiral_pressure_constitutive}
  p_R = \mu_R\left(\Omega - \tfrac{1}{2}\omega\right),
\end{equation}
with the rotational viscosity $\mu_R$ setting the scale of the response, just as the shear viscosity $\mu$
sets the scale of the deviatoric response. The sign of $\mu_R$ is not fixed by this kinematic argument: it
records whether the chiral coupling damps or sustains the relative rotation and, the fluid being active,
either sign is admissible. This freedom is precisely what will separate damped relaxation from active pumping
in the cavity flow of \S\ref{sec:cavity}, through the sign of the parameter $\alpha$.

\subsection{The complete stress tensor and the translation--rotation correspondence}
\label{sec:complete_stress}
 
Collecting the symmetric Newtonian stress \eqref{eq:Newton_stress}, the odd stress \eqref{eq:odd_stress}, and
the constitutive form \eqref{eq:chiral_pressure_constitutive} of the chiral pressure, the complete stress
tensor of a kind-II fluid reads
\begin{align}
  \label{eq:full_stress}
  \sigma^{\mathrm{II}}_{ij} ={}&
    \underbrace{\left[-p + \zeta\,(\boldsymbol{\nabla}\bcdot\boldsymbol{u})\right]\delta_{ij}
    + 2\mu\, d_{ij}}_{\text{symmetric (Newtonian)}}
  \nonumber\\[4pt]
    &{}+ \underbrace{2\mu^{\mathrm{odd}}\, \epsilon_{ik}\, d_{kj}}_{\text{symmetric (odd viscosity)}}
      + \underbrace{\left[\mu_R\!\left(\Omega - \tfrac{1}{2}\omega\right) +
      \zeta^{\mathrm{odd}}\,(\boldsymbol{\nabla}\bcdot\boldsymbol{u})\right]\epsilon_{ij}}_{\text{antisymmetric (chiral)}} .
\end{align}
The six contributions are fixed, with no further constitutive freedom, by three inputs: the response is
Newtonian (linear in the velocity gradients and in the spin), the fluid carries two independent rotation
fields $\Omega$ and $\omega$, and its microscopic chirality breaks parity. Parity breaking is what admits the
three odd terms, all generated from the Newtonian stress by the single rotation
$\delta_{ij}\mapsto\epsilon_{ij}$, $d_{ij}\mapsto\epsilon_{ik}d_{kj}$: the isotropic pressure maps to the
pseudoscalar chiral pressure, the bulk channel to the odd bulk viscosity
$\zeta^{\mathrm{odd}}(\boldsymbol{\nabla}\bcdot\boldsymbol{u})\,\epsilon_{ij}$ --- silent for incompressible
flow --- and the deviatoric viscous stress to the odd-viscous stress. This one-to-one map between the
Newtonian and the odd sectors is summarised in table~\ref{tab:correspondence}.  Note that
$\epsilon_{ik}d_{kj}$ remains symmetric and traceless whenever $d_{ij}$ is symmetric and traceless, so that
the odd-viscosity term belongs to the symmetric part of $\boldsymbol\sigma^{\mathrm{II}}$; the antisymmetric
channel is spanned by $\epsilon_{ij}$ alone, with pseudoscalar coefficient
$p_R + \zeta^{\mathrm{odd}}(\boldsymbol{\nabla}\bcdot\boldsymbol{u})$, reducing to the chiral pressure for
incompressible flow. For the sake of illustration, we have represented in Figure~\ref{fig:stress_actions} and
Figure~\ref{fig:stress_actions_normal} the different types of mechanical actions present in the odd stress.
 
This confinement is not accidental but algebraic: in two dimensions the rotation exchanges the isotropic and
antisymmetric channels of a tensor,
$[\epsilon\!\cdot\!T]^{\mathrm{anti}}_{ij} = \tfrac{1}{2}(\mathrm{tr}\,T)\, \epsilon_{ij}$, while mapping the
symmetric-traceless channel onto itself.  The tracelessness of $d_{ij}$ is therefore precisely what keeps the
odd viscosity out of the couple: had the full strain rate $e_{ij}$ been used instead \citep[as in previous
bibliography, see for instance the work by][]{Banerjee2017}, the compressible trace
would leak the spurious antisymmetric stress
$\mu^{\mathrm{odd}}(\boldsymbol{\nabla}\!\cdot\!\boldsymbol{u})\,\epsilon_{ij}$ --- an artificial chiral
pressure --- into the angular-momentum balance.

\begin{table}
  \begin{center}
  \begin{tabular}{lcl}
    Newtonian (parity-even) sector & $\longrightarrow$ & Odd (parity-odd) sector \\[3pt]
    $\delta_{ij}$ (identity)           & $\longrightarrow$ & $\epsilon_{ij}$ (rotation) \\
    $p$ (pressure, scalar)             & $\longrightarrow$ & $p_R = \mu_R(\Omega-\tfrac{1}{2}\omega)$ (chiral pressure, pseudoscalar) \\
    $\zeta$ (bulk viscosity)           & $\longrightarrow$ & $\zeta^{\mathrm{odd}}$ (odd bulk viscosity, pseudoscalar) \\
    $d_{ij}$ (deviatoric strain rate)  & $\longrightarrow$ & $\epsilon_{ik}d_{kj}$ (rotated deviatoric strain rate) \\
    $\mu$ (shear viscosity)            & $\longrightarrow$ & $\mu^{\mathrm{odd}}$ (odd viscosity, pseudoscalar) \\
  \end{tabular}
  \caption{The translation--rotation correspondence that generates the odd
    stress. Each element of the symmetric Newtonian stress $\boldsymbol\sigma^{\mathrm{I}}$
    maps to its counterpart in the odd (parity-breaking) sector under the
    $90^{\circ}$ rotation represented by the Levi-Civita tensor $\epsilon_{ij}$ of
    \eqref{eq:levi_civita_2D}. The isotropic, bulk and deviatoric channels are mapped
    independently, producing the chiral pressure $p_R$, the odd bulk viscosity $\zeta^{\mathrm{odd}}$ and
    the odd viscosity $\mu^{\mathrm{odd}}$, respectively.}
  \label{tab:correspondence}
  \end{center}
\end{table}

\begin{figure}
  \centering
 
\begin{tikzpicture}[>=Stealth, line cap=round, line join=round,
    sq/.style={draw=black, thick},
    def/.style={draw=drv, dashed, thick},
    drvarr/.style={->, drv, very thick},
    respP/.style={->, newtblue, line width=1.4pt},
    respN/.style={->, oddred,   line width=1.4pt},
    arc/.style={->, drv, thick},
    frm/.style={font=\footnotesize},
    sml/.style={font=\scriptsize},
    lbl/.style={font=\small}, scale=0.80, transform shape]
 
\def\a{0.85}
\def\off{0.16}
\def\sh{0.30}
\def\rh{0.55}
\def\rw{0.60}
\def\arint{0.28}
\def\arext{0.55}
\def\dx{3.5}          
\def\dy{3.15}         
\def\xA{0}
\def\xB{\dx}
\def\xC{2*\dx}
\def\xD{3*\dx}
\def\yA{0}            
\def\yB{-\dy}         
\def\yC{-2*\dy}       
 
\draw[gridcol, thin] (-2.15, {\yA+1.55}) -- ({\xD+1.6}, {\yA+1.55});
\draw[gridcol, thin] (-2.15, {(\yA+\yB)/2-0.15}) -- ({\xD+1.6}, {(\yA+\yB)/2-0.15});
\draw[gridcol, thin] (-2.15, {(\yB+\yC)/2-0.15}) -- ({\xD+1.6}, {(\yB+\yC)/2-0.15});
\draw[gridcol, thin] (-2.15, {\yC-1.95}) -- ({\xD+1.6}, {\yC-1.95});
\draw[gridcol, thin] (-2.15, {\yA+1.55}) -- (-2.15, {\yC-1.95});
\draw[gridcol, thin] ({\xA+\dx/2}, {\yA+1.55}) -- ({\xA+\dx/2}, {\yC-1.95});
\draw[gridcol, thin] ({\xB+\dx/2}, {\yA+1.55}) -- ({\xB+\dx/2}, {\yC-1.95});
\draw[gridcol, thin] ({\xC+\dx/2}, {\yA+1.55}) -- ({\xC+\dx/2}, {\yC-1.95});
 
\node[lbl, align=center] at (-3.05, \yA) {\textbf{Action}};
\node[lbl, align=center] at (-3.05, \yB) {\textbf{Result}\\[1pt] \footnotesize coeff. $>0$};
\node[lbl, align=center] at (-3.05, \yC) {\textbf{Result}\\[1pt] \footnotesize coeff. $<0$};
 
\begin{scope}[shift={(-3.5, {\yA+0.95})}]
  \draw[->] (0,0) -- (0.5,0) node[right, sml] {$x$};
  \draw[->] (0,0) -- (0,0.5) node[above, sml] {$y$};
\end{scope}
 
\node[sml, drv] at (\xA, {\yA+1.20}) {shear strain};
\begin{scope}[shift={(\xA, \yA)}]
  \draw[sq] (-\a,-\a) rectangle (\a,\a);
  \draw[def] (-\a-\sh,-\a) -- (\a-\sh,-\a) -- (\a+\sh,\a) -- (-\a+\sh,\a) -- cycle;
  \draw[drvarr] (-0.42,\a+0.30) -- (0.52,\a+0.30);
  \draw[drvarr] (0.42,-\a-0.30) -- (-0.52,-\a-0.30);
  \node[sml] at (0,-\a-0.62) {$\sigma^{\mathrm{I}}_{xy}=2\mu\,d_{xy}$};
\end{scope}
\begin{scope}[shift={(\xA, \yB)}]
  \draw[sq] (-\a,-\a) rectangle (\a,\a);
  \draw[respP] (\a+\off,-0.55) -- (\a+\off,0.55);
  \draw[respP] (-0.55,\a+\off) -- (0.55,\a+\off);
  \draw[respP] (-\a-\off,0.55) -- (-\a-\off,-0.55);
  \draw[respP] (0.55,-\a-\off) -- (-0.55,-\a-\off);
  \node[sml, drv] at (0,-\a-0.62) {$\mu>0$};
\end{scope}
\begin{scope}[shift={(\xA, \yC)}]
  \draw[gridcol, thick, dashed] (-\a,-\a) rectangle (\a,\a);
  \draw[gridcol, thick] (-\a,-\a) -- (\a,\a);
  \draw[gridcol, thick] (-\a,\a) -- (\a,-\a);
  \node[sml, drv, align=center] at (0,-\a-0.72) {$\mu<0$: forbidden\\by thermodynamics};
\end{scope}
 
\node[sml, drv] at (\xB, {\yA+1.20}) {normal strain};
\begin{scope}[shift={(\xB, \yA)}]
  \draw[sq] (-\a,-\a) rectangle (\a,\a);
  \draw[def] (-\rw,-\rh) rectangle (\rw,\rh);
  \draw[drvarr] (0,\a+0.10) -- (0,\a+0.55);
  \draw[drvarr] (0,-\a-0.10) -- (0,-\a-0.55);
  \draw[drvarr] (\a+0.55,0) -- (\a+0.10,0);
  \draw[drvarr] (-\a-0.55,0) -- (-\a-0.10,0);
  \node[sml] at (0,-\a-0.72) {$\sigma^{\mathrm{odd,sym}}_{xy}=2\mu^{\mathrm{odd}}d_{yy}$};
\end{scope}
\begin{scope}[shift={(\xB, \yB)}]
  \draw[sq] (-\a,-\a) rectangle (\a,\a);
  \draw[respP] (\a+\off,-0.55) -- (\a+\off,0.55);
  \draw[respP] (-0.55,\a+\off) -- (0.55,\a+\off);
  \draw[respP] (-\a-\off,0.55) -- (-\a-\off,-0.55);
  \draw[respP] (0.55,-\a-\off) -- (-0.55,-\a-\off);
  \node[sml, drv] at (0,0) {as col.\ 1};
  \node[sml, drv] at (0,-\a-0.62) {$\mu^{\mathrm{odd}}>0$};
\end{scope}
\begin{scope}[shift={(\xB, \yC)}]
  \draw[sq] (-\a,-\a) rectangle (\a,\a);
  \draw[respN] (\a+\off,0.55) -- (\a+\off,-0.55);
  \draw[respN] (0.55,\a+\off) -- (-0.55,\a+\off);
  \draw[respN] (-\a-\off,-0.55) -- (-\a-\off,0.55);
  \draw[respN] (-0.55,-\a-\off) -- (0.55,-\a-\off);
  \node[sml, drv] at (0,-\a-0.62) {$\mu^{\mathrm{odd}}<0$};
\end{scope}
 
\node[sml, drv, align=center] at (\xC, {\yA+1.20}) {spin--vorticity\\[-1pt] mismatch};
\begin{scope}[shift={(\xC, \yA)}]
  \draw[sq] (-\a,-\a) rectangle (\a,\a);
  \draw[arc, oddred, thick] ({\arext*cos(-20)},{\arext*sin(-20)})
    arc[start angle=-20, end angle=260, radius=\arext];
  \node[sml, oddred] at (\arext+0.26, -0.05) {$\Omega$};
  \draw[arc, newtblue, thick] ({\arint*cos(-30)},{\arint*sin(-30)})
    arc[start angle=-30, end angle=250, radius=\arint];
  \node[sml, newtblue] at (0.0,0) {$\tfrac{\omega}{2}$};
  \node[sml] at (0,-\a-0.62) {$p_R=\mu_R(\Omega-\tfrac{\omega}{2})$};
\end{scope}
\begin{scope}[shift={(\xC, \yB)}]
  \draw[sq] (-\a,-\a) rectangle (\a,\a);
  \draw[respP] (-0.55,\a+\off) -- (0.55,\a+\off);
  \draw[respP] (\a+\off,0.55) -- (\a+\off,-0.55);
  \draw[respP] (0.55,-\a-\off) -- (-0.55,-\a-\off);
  \draw[respP] (-\a-\off,-0.55) -- (-\a-\off,0.55);
  \node[sml, newtblue] at (0,0) {(CW)};
  \node[sml, drv] at (0,-\a-0.62) {$\mu_R>0$};
\end{scope}
\begin{scope}[shift={(\xC, \yC)}]
  \draw[sq] (-\a,-\a) rectangle (\a,\a);
  \draw[respN] (0.55,\a+\off) -- (-0.55,\a+\off);
  \draw[respN] (\a+\off,-0.55) -- (\a+\off,0.55);
  \draw[respN] (-0.55,-\a-\off) -- (0.55,-\a-\off);
  \draw[respN] (-\a-\off,0.55) -- (-\a-\off,-0.55);
  \node[sml, oddred] at (0,0) {(CCW)};
  \node[sml, drv] at (0,-\a-0.62) {$\mu_R<0$};
\end{scope}
 
\node[sml, drv] at (\xD, {\yA+1.20}) {dilation};
\begin{scope}[shift={(\xD, \yA)}]
  \draw[sq] (-\a,-\a) rectangle (\a,\a);
  \draw[def] (-\a-\sh,-\a-\sh) rectangle (\a+\sh,\a+\sh);
  \draw[drvarr] (0,\a+0.10) -- (0,\a+0.55);
  \draw[drvarr] (0,-\a-0.10) -- (0,-\a-0.55);
  \draw[drvarr] (\a+0.10,0) -- (\a+0.55,0);
  \draw[drvarr] (-\a-0.10,0) -- (-\a-0.55,0);
  \node[sml] at (0,-\a-0.72) {$\sigma^{\mathrm{odd,anti}}_{xy}=\zeta^{\mathrm{odd}}\boldsymbol{\nabla}\bcdot\boldsymbol{u}$};
\end{scope}
\begin{scope}[shift={(\xD, \yB)}]
  \draw[sq] (-\a,-\a) rectangle (\a,\a);
  \draw[respP] (-0.55,\a+\off) -- (0.55,\a+\off);
  \draw[respP] (\a+\off,0.55) -- (\a+\off,-0.55);
  \draw[respP] (0.55,-\a-\off) -- (-0.55,-\a-\off);
  \draw[respP] (-\a-\off,-0.55) -- (-\a-\off,0.55);
  \node[sml, newtblue] at (0,0) {(CW)};
  \node[sml, drv] at (0,-\a-0.62) {$\zeta^{\mathrm{odd}}>0$};
\end{scope}
\begin{scope}[shift={(\xD, \yC)}]
  \draw[sq] (-\a,-\a) rectangle (\a,\a);
  \draw[respN] (0.55,\a+\off) -- (-0.55,\a+\off);
  \draw[respN] (\a+\off,-0.55) -- (\a+\off,0.55);
  \draw[respN] (-0.55,-\a-\off) -- (0.55,-\a-\off);
  \draw[respN] (-\a-\off,0.55) -- (-\a-\off,-0.55);
  \node[sml, oddred] at (0,0) {(CCW)};
  \node[sml, drv] at (0,-\a-0.62) {$\zeta^{\mathrm{odd}}<0$};
\end{scope}
 
\end{tikzpicture}
\caption{Mechanical action of the channels of the shear component $\sigma^{\mathrm{II}}_{xy}$ in a
  kind-II fluid. Arrows in the Result rows denote the tractions exerted \textit{on} the fluid element by its
  surroundings ($t_j = \sigma_{jl}n_l$, with $\boldsymbol{n}$ the outward normal); each column shows one
  channel, with the response for either sign of the corresponding transport coefficient. From left to right:
  the Newtonian shear stress $\sigma^{\mathrm{I}}_{xy}$ --- shear begets shear; the symmetric odd stress
  $\sigma^{\mathrm{odd,sym}}_{xy}$ --- normal strain begets shear traction, the $90^{\circ}$ rotation of the
  deviatoric channel by $\epsilon_{ij}$; and the two channels that produce a net couple, the chiral pressure
  $p_R$ --- spin--vorticity mismatch begets coherently circulating tractions, for $p_R>0$ clockwise, their
  moment being the sink $-2p_R$ of the spin balance
  \eqref{eq:angular_momentum_convective_form} that drains the excess spin --- and the odd bulk viscosity
  $\zeta^{\mathrm{odd}}$ --- dilation likewise begets circulating tractions, this channel being silent in the
  incompressible flows considered in this work. The odd coefficients $\mu^{\mathrm{odd}}$ and
  $\zeta^{\mathrm{odd}}$, being non-dissipative, may take either sign, whereas $\mu_R<0$ is forbidden for a
  passive fluid on the same thermodynamic grounds as $\mu<0$ --- the dissipation of the antisymmetric channel
  being proportional to $\mu_R(\Omega-\omega/2)^2$ \citep{Condiff_1964} --- and is accessible only as an
  effective coefficient sustained by activity (\S\ref{sec:cavity_interpretation}). The figure is drawn for
  $\Omega > \omega/2$ (co-rotating branch); reversing the mismatch reverses the sense of circulation.
  \label{fig:stress_actions}}
\end{figure}
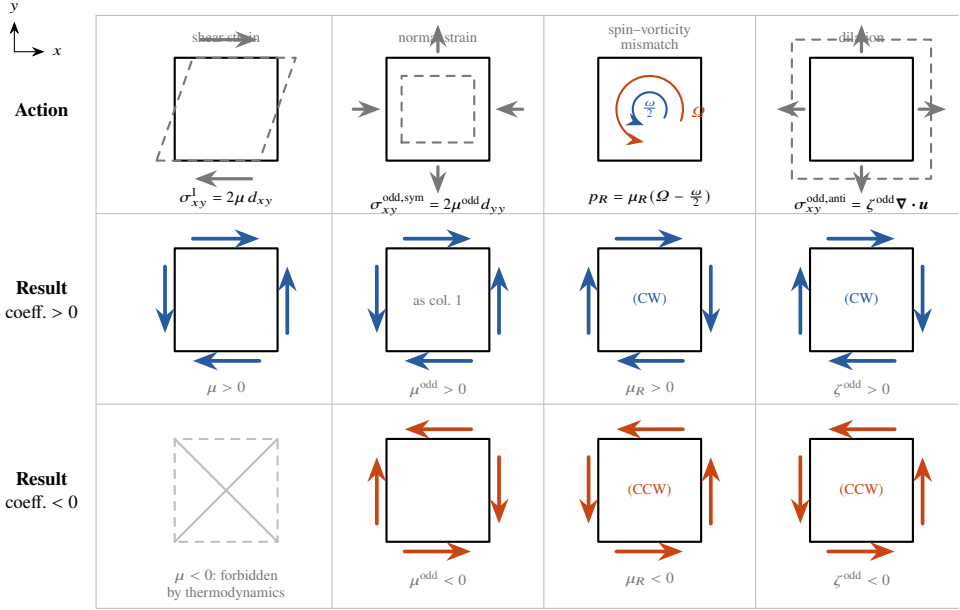

\begin{figure}
  \centering
 
\begin{tikzpicture}[>=Stealth, line cap=round, line join=round,
    sq/.style={draw=black, thick},
    def/.style={draw=drv, dashed, thick},
    drvarr/.style={->, drv, very thick},
    respP/.style={->, newtblue, line width=1.4pt},
    respN/.style={->, oddred,   line width=1.4pt},
    frm/.style={font=\footnotesize},
    lbl/.style={font=\small},scale=0.825, transform shape]
 
\def\a{0.85}
\def\off{0.16}
\def\sh{0.30}
\def\dx{4.5}
 
\draw[gridcol, thin] (-2.2, 2.0) -- (12.0, 2.0);
\draw[gridcol, thin] (-2.2, -2.0) -- (12.0, -2.0);
\draw[gridcol, thin] ({\dx-\a-1.7}, 2.0) -- ({\dx-\a-1.7}, -2.0);
\draw[gridcol, thin] ({2*\dx-\a-1.7}, 2.0) -- ({2*\dx-\a-1.7}, -2.0);
 
\node[lbl] at (-2.0, 2.9) {$\sigma^{\mathrm{II}}_{xx}$};
 
\node[lbl] at (0, 2.35) {\textbf{Action}};
\node[lbl] at (\dx, 2.35) {\textbf{Result}};
\node[lbl] at (2*\dx, 2.35) {\textbf{Result}};
\node[frm, drv] at (\dx, 1.7) {$\mu^{\mathrm{odd}}>0$};
\node[frm, drv] at (2*\dx, 1.7) {$\mu^{\mathrm{odd}}<0$};
 
\begin{scope}[shift={(-2.0, 0.6)}]
  \draw[->] (0,0) -- (0.55,0) node[right, frm] {$x$};
  \draw[->] (0,0) -- (0,0.55) node[above, frm] {$y$};
\end{scope}
 
\begin{scope}[shift={(0, 0)}]
  \draw[sq] (-\a,-\a) rectangle (\a,\a);
  \draw[def] (-\a-\sh,-\a) -- (\a-\sh,-\a) -- (\a+\sh,\a) -- (-\a+\sh,\a) -- cycle;
  \draw[drvarr] (-0.42,\a+0.30) -- (0.52,\a+0.30);
  \draw[drvarr] (0.42,-\a-0.30) -- (-0.52,-\a-0.30);
  \node[frm] at (0,-\a-0.85) {$\sigma^{\mathrm{odd}}_{xx} = 2\mu^{\mathrm{odd}}\,d_{xy}$};
  \node[frm, drv] at (0,-\a+2.57) {shear strain};
\end{scope}
 
\begin{scope}[shift={(\dx, 0)}]
  \draw[sq] (-\a,-\a) rectangle (\a,\a);
  \draw[respP] (\a+0.10,0) -- (\a+0.55,0);
  \draw[respP] (-\a-0.10,0) -- (-\a-0.55,0);
  \node[frm, newtblue] at (\a+0.72,0.82) {$\sigma^{\mathrm{odd}}_{xx}$};
\end{scope}
 
\begin{scope}[shift={(2*\dx, 0)}]
  \draw[sq] (-\a,-\a) rectangle (\a,\a);
  \draw[respN] (\a+0.55,0) -- (\a+0.10,0);
  \draw[respN] (-\a-0.55,0) -- (-\a-0.10,0);
  \node[frm, oddred] at (\a+0.72,0.82) {$\sigma^{\mathrm{odd}}_{xx}$};
\end{scope}
 
\end{tikzpicture}
\caption{Mechanical action of the odd channel of the diagonal component
  $\sigma^{\mathrm{odd}}_{xx} = 2\mu^{\mathrm{odd}}d_{xy}$, with the same traction convention as in
  figure~\ref{fig:stress_actions}: shear strain produces normal traction, completing the $90^{\circ}$
  correspondence --- shear $\to$ normal here mirrors normal $\to$ shear there. Together, the two figures
  display the full action of the parity-breaking operator $\epsilon$ on the symmetric-traceless sector of the
  Newtonian stress.
  \label{fig:stress_actions_normal}}
\end{figure}
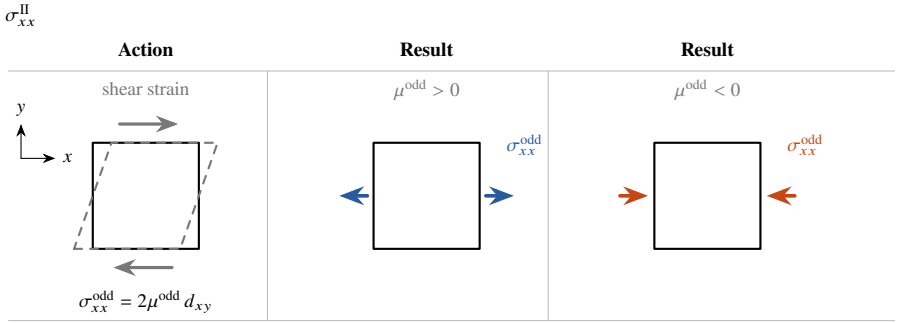

\subsection{The spin flux}
\label{sec:spin_flux}

We finally determine the constitutive form of the spin flux $\boldsymbol{C}$ introduced in the angular
momentum balance \eqref{eq:L_balance_chiral}. Physically, $\boldsymbol{C}$ accounts for the transport of
internal (particle) angular momentum across the boundary of a fluid element through direct spin exchange
between neighbouring particles. It is therefore the rotational analogue of the momentum flux $\sigma$, and
coincides with the couple stress of the hydrodynamics of structured fluids \citep{Condiff_1964}.

The constitutive form of $\boldsymbol{C}$ follows from the same principles that fixed the stress tensor in
\S\ref{sec:odd_stress}. In two dimensions the spin field is a pseudoscalar, so its flux $\boldsymbol{C}$ must
be a vector of odd parity. At Navier--Stokes order, constitutive relations are linear in the first spatial
gradients of the hydrodynamic fields \citep{Kim_1998}. The only available gradient with the required
tensorial character is $\partial_l\Omega$: the vorticity gradient $\partial_l\omega$, being a second
derivative of the velocity field, belongs to Burnett order \cite{Burnett_1935,Brush_1972} and is therefore
excluded from the present description.\footnote{Since the substrate drag breaks Galilean invariance, a
  gradient-free term proportional to $\epsilon_{lk}u_k$ is also allowed by symmetry. Its divergence, however,
  is proportional to the vorticity and merely renormalises the coefficient of $\omega$ in the spin balance, so
  it may be absorbed into $\mu_R$ without loss of generality; a term proportional to $u_l$ would contribute
  only for compressible flow.}  Two invariant combinations can be formed from $\partial_l\Omega$ --- the
gradient itself and its image under the rotation operation \eqref{eq:levi_civita_2D}, in analogy to our
procedure for the stress tensor in \S\ref{sec:odd_stress} --- and hence the most general spin flux at this
order is
\begin{equation}
  \label{eq:C_constitutive}
  C_l = \kappa\, \partial_l \Omega
  + \kappa^\mathrm{odd}\, \epsilon_{lk}\, \partial_k \Omega ,
\end{equation} where $\kappa$, $\kappa^{\mathrm{odd}}$ are two new transport coefficients that we call spin
viscosity and odd spin viscosity respectively.
The first term, with spin viscosity $\kappa \geq 0$, describes the diffusive transport of internal angular
momentum and is inherent in any structured fluid, chiral or not \citep{Condiff_1964}. The second term is its
chiral counterpart: since $\epsilon_{lk}\partial_k\Omega$ is a true vector, the coefficient
$\kappa^\mathrm{odd}$ must be a pseudoscalar, admissible only when parity is broken --- for instance, by
means of a
persistent active torque --- exactly as the odd viscosity $\mu^\mathrm{odd}$ descends from
the Newtonian $\mu$ through the same rotation operation.

\section{Incompressible flow for fluids of the kind II}
\label{sec:incompressible}

We analyse now the properties of incompressible flow ($\boldsymbol\nabla\cdot\boldsymbol{u} = 0$) with
constant transport coefficients for fluids of the kind II.

First, the bulk viscosity terms both will vanish since they are proportional to
$\boldsymbol\nabla\cdot\boldsymbol{u}$. Next, another important result is that for incompressible flow 
$\sigma^* = -p_R$, which we use for the next sections. Remarkably, for incompressible flow with constant spin viscosities, the chiral component of the spin flux is
identically divergence-free. We take the divergence in \eqref{eq:C_constitutive}
\begin{equation}
  \label{eq:divC}
  \boldsymbol{\nabla}\bcdot\boldsymbol{C}
  = \kappa\,\nabla^2\Omega
  + \kappa^\mathrm{odd}\,\epsilon_{lk}\,\partial_l\partial_k\Omega
  = \kappa\,\nabla^2\Omega ,
\end{equation}
since the full contraction of the antisymmetric tensor $\epsilon_{lk}$ with the symmetric tensor
$\partial_l \partial_k \Omega$ vanishes. Hence $\kappa^\mathrm{odd}$, like the odd viscosity
$\mu^\mathrm{odd}$ in the vorticity balance, is silent in the bulk equations: in two-dimensional confined
flow, chirality acts on the interior dynamics through the chiral pressure alone. Neither odd coefficient
alters the velocity, vorticity or spin fields: $\mu^\mathrm{odd}$ is recovered from the pressure field, which
it shifts by the amount \eqref{eq:pressure_shift}, and from the tractions exerted on the boundaries, while
$\kappa^\mathrm{odd}$ survives only in the couples transmitted to them. This structural parallelism
between $\{\mu, \mu^\mathrm{odd}\}$ and $\{\kappa, \kappa^\mathrm{odd}\}$ is, to our knowledge, not emphasised
in the previous literature.

Next, we show that the remaining contribution to $\boldsymbol\nabla\cdot\boldsymbol{C}$ in \eqref{eq:divC} is
negligible under incompressible flow. In effect, with the closure \eqref{eq:divC}, and the constitutive form
\eqref{eq:chiral_pressure_constitutive} of the chiral pressure, the spin balance
\eqref{eq:angular_momentum_convective_form} for steady states reads

\begin{equation}
  \label{eq:spin_balance_full}
  \kappa\,\nabla^2\Omega - 2\mu_R\Omega + \mu_R\,\omega + \tau_0 = 0 .
\end{equation}

The relative weight of spin diffusion is measured by the spin screening length
\begin{equation}
  \label{eq:spin_screening_length}
  \ell_s = \sqrt{\frac{\kappa}{2\mu_R }} .
\end{equation}
On dimensional grounds $\kappa \sim \mu_R\, a^2$, with $a$ the particle radius --- an estimate consistent
with the microscopic expressions for the couple stress obtained by coarse-graining the equations of motion of
torque-driven particles \citep{Klymko_2017} --- so that $\ell_s \lesssim a$:
the screening length is microscopic. Thus, for a cavity of linear size $L \gg \ell_s$, the diffusive term in
\eqref{eq:spin_balance_full} is negligible everywhere except within boundary layers of thickness $O(\ell_s)$,
which lie below the resolution of the continuum description itself and for this reason we can take
$\boldsymbol\nabla\cdot\boldsymbol{C} = \boldsymbol{0}$ in the bulk. 

Thus, for incompressible flow and constant transport coefficients, the system of equations
\eqref{eq:mass_convective_form}--\eqref{eq:angular_momentum_convective_form} reduces to

\begin{align}
  &\frac{\mathrm{D}\rho}{\mathrm{D}t} =0 \label{eq:incompressible_density}\\
  \rho &\frac{\mathrm{D}\boldsymbol{u}}{\mathrm{D}t} = \boldsymbol{f} -\nabla p + \nabla^2 (\mu\boldsymbol{u})
    +  \nabla^2 (\mu^{\mathrm{odd}}\boldsymbol{u^*}) + \nabla^* p_R, \label{eq:incompressible_moment} \\
  \rho I &\frac{\mathrm{D}\Omega}{\mathrm{D}t} = -2 p_R + \tau_0, \label{eq:incompressible_angular_moment}
\end{align} where $u^*_i \equiv \epsilon_{ij} u_j$, $\nabla_i^* \equiv \epsilon_{ij}\nabla_j$ and we have taken into account the
approximation of negligible spin flux in the volume of the fluid, as explained in \S\ref{sec:spin_flux}. We
also took into account
the simplifications for the stress tensor $\boldsymbol\sigma^{\mathrm{II}}$ under the incompressibility condition, since for
$\boldsymbol\nabla\bcdot\boldsymbol{u} = \boldsymbol{0}$ we obtain $d_{ij} \to e_{ij}$. For instance, the full
derivation for the terms $\boldsymbol{\sigma}_{xj}^{II}$, present in the X component of the momentum balance
\eqref{eq:incompressible_moment}, yields:

\par\medskip

\begin{itemize}
  \item for the isotropic part of $\sigma_{xj}^I$: $\partial_j(-p\,\delta_{xj}) = -\partial_x p$
\item for the deviatoric part of $\sigma_{xj}^I$:
  $\partial_{j}(2\mu\, e_{xj}) = \mu\,\partial_j(\partial_j u_x+ \partial_x u_j) = \mu\, \partial_j^2u_x +
  \mu\,\partial_x (\partial_ju_j) = \mu\nabla^2u_x$.

\item for the deviatoric part of $\boldsymbol{\sigma}^{\mathrm{odd}}_{xj}$:  $2\mu^{\mathrm{odd}}\partial_j(\epsilon_{xk}\,e_{kj}) =
2\mu^{\mathrm{odd}}\partial_j(\epsilon_{xy}\,e_{yj}) = \mu^{\mathrm{odd}}\partial_j (\partial_ju_y + \partial_y
u_j) = \mu^{\mathrm{odd}}\partial_j^2u_y + \mu^{\mathrm{odd}}\partial_y(\partial_j u_j) =
\mu^{\mathrm{odd}}\nabla^2u_y$.
\item for the isotropic part of $\boldsymbol\sigma^{\mathrm{odd}}$:
  $\partial_j(\epsilon_{xj}\,p_R) = \epsilon_{xy}\partial_yp_R = \partial_yp_R$.
\end{itemize}

\par\medskip

And analogously for the terms from $\partial_j\boldsymbol{\sigma}^{II}_{yj} $. Note that now the spin balance
under steady state therefore reduces to the local (algebraic) relation
\begin{equation} 
  \label{eq:spin_closure_pre}
  2\mu_R\Omega = \tau_0 + \mu_R\,\omega ,
\end{equation}
which means that for incompressible steady flow the spin field is slaved to the vorticity field.

\section{Simple steady base states for chiral fluids (kind II-a)}
\label{sec:base_chiral}


In this section, we analyze two of the simplest cases of steady states for a chiral fluid whose particles are
subject to active rotation. This classifies it as a kind II-a fluid. In particular, we will work under a model
for which the particles are subject to an intrinsic active rotation torque $\tau_a$ and no external field, so
that $\tau_b = 0$ and $\tau_0 = \tau_a + \tau_{\mathrm{drag}} = \tau_a - \Gamma^\Omega\Omega$ in the notation
of \S\ref{sec:kind_II}. The volumetric force reduces to $\boldsymbol{f} = -\Gamma\boldsymbol{u}$, the
translational counterpart of the rotational drag. Incorporating these specific forms of body couple and volume
force into equations \eqref{eq:incompressible_density}--\eqref{eq:incompressible_angular_moment} we obtain

\begin{align}
  &\frac{\mathrm{D}\rho}{\mathrm{D}t} =0 \label{eq:chiral_density}\\
  \rho &\frac{\mathrm{D}\boldsymbol{u}}{\mathrm{D}t} = -\Gamma\boldsymbol{u} -\nabla p +
         \mu \nabla^2 \boldsymbol{u}
         +  \mu^{\mathrm{odd}} \nabla^2 \boldsymbol{u^*} + \nabla^* p_R, \label{eq:chiral_moment} \\
  \rho I &\frac{\mathrm{D}\Omega}{\mathrm{D}t} = - 2 \mu_R\left(\Omega - \frac{\omega}{2}\right)-
           \Gamma^\Omega \Omega + \tau_a,
         \label{eq:chiral_angular_moment}
\end{align}

The microscopic origin of the couple in kind II-a fixes its dependence on the density. Being a couple
per unit volume, $\tau_0$ is trivially extensive: if each particle receives a material torque $\tau_p$ from
its internal mechanism, then $\tau_a = n\,\tau_p$, with $n$ the number density, at every density. The
non-trivial statement concerns the intensive quantities: in the dilute regime $\tau_p$ --- and with it the
free spinning rate $\Omega_\mathrm{free} = \tau_a/\Gamma^\Omega = \tau_p/\gamma_R$, since the rotational drag
density $\Gamma^\Omega = n\,\gamma_R$ carries the same trivial factor --- is independent of $n$, while the
collective renormalisation of the couple and friction densities arises from pair interactions and enters at
order $n^2$, the scaling reported for the collisional rotational friction of actively spinning grains
\citep{Han_2021}. The steady base states studied in this work have uniform density, so that $\tau_a$ is
spatially constant and thus $\nabla^2\tau_a = 0$. In inhomogeneous states,
$\boldsymbol{\nabla} n \neq \boldsymbol{0}$ converts $\tau_0(\boldsymbol{r})$ into a spatial source coupling
the density and spin fields; the phenomenology of such states, exemplified by rotation-induced phase
separation \citep{Digregorio_2025}, lies beyond the homogeneous base states addressed here.

Note that $\mu^{\mathrm{odd}}\nabla^2\boldsymbol{u}^* =
\mu^{\mathrm{odd}}\boldsymbol{\nabla}\omega$ which, used in \eqref{eq:chiral_moment}, leads to a renormalised
effective pressure $p^* = p -\mu^{\mathrm{odd}}\omega$: in the incompressible bulk the odd viscosity enters
only through the pressure and drives no flow.  Once parity is broken (due to the existence of a non-vanishing
fluid angular momentum), the isotropic antisymmetric channel (the chiral pressure) and the rotated deviatoric
channel (the odd viscosity), together with their spin-flux counterparts $\{\kappa, \kappa^\mathrm{odd}\}$,
exhaust the admissible linear response in the trace-free sectors of the stress; no further freedom remains
there at Navier--Stokes order, for incompressible flow. The single remaining structure allowed by symmetry ---
a reactive pressure--rotation coupling $\propto \omega\,\delta_{ij}$, recently measured in chiral granular
gases \citep{Han_2021} --- exerts a pure-gradient force $\propto \boldsymbol{\nabla}\omega$, absorbed into the
pressure for incompressible flow exactly as the odd viscous force of \S\ref{sec:odd_rotated_force}, and plays
no role in what follows. For this reason, our framework is exact at this order for the flows considered here
and yields systematically the set of base steady states in the kind-II fluids. As stated above, we focus here
on kind II-a fluids.

\subsection{Hydrostatic states and the chiral complex potential}
\label{sec:hydrostatic}

We first ask which states of the chiral fluid can remain at rest. In a classical fluid free of body forces,
mechanical equilibrium requires a uniform pressure; the chiral fluid, as we now show, admits a far richer
family of quiescent states, organised by an unexpected mathematical structure. Setting
$\boldsymbol{u}=\boldsymbol{0}$ (hence $\omega = 0$) and $\partial_t = 0$ in
\eqref{eq:chiral_moment}--\eqref{eq:chiral_angular_moment}, the balance equations reduce to
\begin{align}
  \boldsymbol\nabla p &= \boldsymbol{\nabla}^* p_R ,
  \label{eq:hydrostatic_moment}\\
  \left(2\mu_R + \Gamma^\Omega\right)\Omega &= \tau_a ,
  \label{eq:hydrostatic_angular_moment}
\end{align}
while the mass balance is trivially satisfied. The spin balance \eqref{eq:hydrostatic_angular_moment} is
algebraic and local, so the resting fluid carries the spin field $\Omega = \tau_a/(2\mu_R+\Gamma^\Omega)$
and, by \eqref{eq:chiral_pressure_constitutive} with $\omega = 0$, the chiral pressure
\begin{equation}
  \label{eq:pR_hydrostatic}
  p_R = \beta\,\tau_a , \qquad \beta \equiv \frac{\mu_R}{2\mu_R + \Gamma^\Omega} :
\end{equation}
at rest, the chiral pressure is simply proportional to the applied torque density. Here, we defined the
\textit{entrainment coefficient} $\beta$.

The structure of \eqref{eq:hydrostatic_moment} is, however, qualitatively different from that of classical
hydrostatics: the mechanical pressure is balanced not by the gradient of the chiral pressure but by its
$90^{\circ}$ rotation $\boldsymbol\nabla^*$. Written in components,
\begin{equation}
  \label{eq:CR_pair}
  \frac{\partial p}{\partial x} = \frac{\partial p_R}{\partial y} ,
  \qquad
  \frac{\partial p}{\partial y} = -\,\frac{\partial p_R}{\partial x} ,
\end{equation}
which are precisely the Cauchy--Riemann equations for the pair $(p,\,p_R)$. Equivalently, the complex field
\begin{equation}
  \label{eq:chiral_complex_potential}
  \chi(z) \equiv p + i\,p_R , \qquad z = x + i y ,
\end{equation}
is a holomorphic function of $z$, which we term the \textit{chiral complex potential}. The orientation of
the pair is not accidental: the sign convention fixed in \S\ref{sec:odd_stress} --- the chiral pressure
entering the stress as $+p_R\,\epsilon_{ij}$, i.e.\ $p_R \equiv -\sigma^*$ --- is precisely the one that
renders $\chi$ holomorphic rather than anti-holomorphic (a function of $\bar z$). The hydrostatic
states of a two-dimensional chiral fluid are thus in one-to-one correspondence with analytic functions, in
exact formal analogy with the complex potential $w(z) = \phi + i\psi$ of classical two-dimensional
irrotational flow \citep{Batchelor_1967}. The dictionary between the two theories is worth spelling out. In
the ideal-flow case the real and imaginary parts of $w$ are the velocity potential and the streamfunction,
and the Cauchy--Riemann equations express simultaneously the irrotationality and the incompressibility of
the flow; here the real and imaginary parts of $\chi$ are the mechanical and the chiral pressures, and the
Cauchy--Riemann equations express the two components of mechanical equilibrium under the transverse chiral
force. In both cases the two real fields form a conjugate harmonic pair, and their level curves are mutually
orthogonal families: just as streamlines cross equipotential lines at right angles, the isobars of a resting
chiral fluid cross the level lines of the chiral pressure at right angles. This orthogonality is the
hydrostatic fingerprint of chirality --- a stress of pressure-like magnitude that pushes not along its own
gradient but perpendicular to it.

Two consequences follow immediately from holomorphy. First, both members of the pair are harmonic,
\begin{equation}
  \nabla^2 p = \nabla^2 p_R = 0 \, ,
  \label{eq:hydrostatic_harmonic_p}
\end{equation}
so that by \eqref{eq:pR_hydrostatic} a quiescent state exists only if the applied torque density is itself
harmonic,
\begin{equation}
  \label{eq:harmonic_condition}
  \nabla^2 \tau_a = 0 .
\end{equation}
A chiral fluid subject to a non-harmonic torque distribution --- a localised patch of activity, for
instance --- cannot remain at rest: flow is not merely possible but obligatory. This observation anticipates
the mechanism at work in the confined flows of \S\ref{sec:cavity}, where precisely such an obstruction
drives the chiral circulation. Second, when \eqref{eq:harmonic_condition} holds, the pressure is recovered
from the chiral pressure by the quadrature
\begin{equation}
  \label{eq:p_quadrature}
  p(\boldsymbol{r}) = p(\boldsymbol{r}_0)
  + \int_{\boldsymbol{r}_0}^{\boldsymbol{r}} \epsilon_{ij}\,\partial_j p_R \,\mathrm{d}l_i ,
\end{equation}
independently of the path in any simply connected domain; $p$ is determined, up to a constant, as the
harmonic conjugate of $p_R$. (In a multiply connected domain, single-valuedness of the pressure requires in
addition that the flux $\oint \partial_n p_R\,\mathrm{d}s$ vanish around every internal boundary, in exact
parallel with the circulation periods of the classical complex potential.)

A further property makes this family special. The algebraic closure used throughout this work was justified
in \S\ref{sec:spin_flux} by the smallness of the spin screening length, $\ell_s \ll L$. For the hydrostatic
family no such approximation is needed: when $\tau_a$ is harmonic, the spin field
$\Omega = \tau_a/(2\mu_R+\Gamma^\Omega)$ is harmonic as well, and the diffusive term $\kappa\nabla^2\Omega$
in \eqref{eq:spin_balance_full} vanishes identically. The quiescent states
\eqref{eq:hydrostatic_moment}--\eqref{eq:pR_hydrostatic} are therefore exact solutions of the full system,
spin diffusion included, for arbitrary $\ell_s/L$.

The simplest member of the family is the homogeneous state generated by a uniform torque: $\tau_a$
constant, $\chi$ reduced to a constant, $p$ uniform and $p_R = \beta\tau_a$ uniform. Although
featureless in the bulk, this state is not stress-free. The traction transmitted across any surface of unit
normal $\boldsymbol{n}$ contains the chiral contribution $p_R\,\epsilon_{ij}n_j$, which is purely tangential
and of uniform magnitude $|p_R|$: the fluid at rest exerts a net torque on its container. For a circular wall
of radius $R$, the torque exerted by the fluid on the wall is
\begin{equation}
  \label{eq:boundary_torque}
  T = 2\pi R^2\, p_R = 2\pi R^2\,\beta\,\tau_a , 
\end{equation}
and the global angular-momentum balance closes exactly: since $\Omega$ is uniform the couple-stress flux
vanishes, and $T$ equals the net couple injected in the bulk,
$\int (\tau_a - \Gamma^\Omega\Omega)\,\mathrm{d}A = 2\pi R^2 \beta\tau_a$. The rotational activity that
survives the rotational drag is disposed of entirely through the walls. This is the cleanest realisation of
the boundary transmission established in \S\ref{sec:incompressible}: a quiescent chiral suspension transmits
a measurable wall torque, and the ratio $T/\tau_a$
provides a flow-free measurement of the coupling $\beta$, and hence of $\mu_R/(2\mu_R+\Gamma^\Omega)$.

The simplest non-trivial member illustrates the geometry of the theory. For a linear activity profile
$\tau_a = \bar{\tau} + G x$ --- harmonic, so that \eqref{eq:harmonic_condition} is satisfied --- the chiral
complex potential is linear, $\chi(z) = (p_0 + i\beta\bar{\tau}) + i\beta G\, z$, and the pressure field reads
\begin{equation}
  \label{eq:linear_example}
  p = p_0 - \beta G\, y .
\end{equation}
An activity gradient along $x$ sustains, with strictly zero flow, a hydrostatic pressure gradient along
$y$. In a classical fluid a hydrostatic pressure gradient requires a parallel body force; in the chiral
fluid it is maintained by a \emph{perpendicular} gradient of activity, the transverse chiral force
$(\boldsymbol{\nabla}p_R)^*$ supplying exactly the balancing push. This rotated response, obtained here at
rest and in closed form, is the static counterpart of the transverse flows analysed in \S\ref{sec:cavity}.

The same construction in polar coordinates $(r,\theta)$ yields the radially symmetric states and, with
them, a concrete illustration of the topological caveat noted below \eqref{eq:p_quadrature}. In polar form
the Cauchy--Riemann pair \eqref{eq:CR_pair} reads
\begin{equation}
  \label{eq:CR_polar}
  \frac{\partial p}{\partial r} = \frac{1}{r}\,\frac{\partial p_R}{\partial \theta} ,
  \qquad
  \frac{1}{r}\,\frac{\partial p}{\partial \theta} = -\,\frac{\partial p_R}{\partial r} .
\end{equation}
If the pressure is to depend on $r$ alone, the second relation forces $p_R$ to depend on $\theta$ alone,
and the first then requires that dependence to be linear: the radial-pressure states form the one-parameter
logarithmic family
\begin{equation}
  \label{eq:log_state}
  \chi(z) = A \log z + \mathrm{const} , \qquad
  p = p_0 + A \ln (r/r_0) , \qquad
  p_R = A\,\theta ,
\end{equation}
that is, $\tau_a = (A/\beta)\,\theta$. In a wedge-shaped domain $0 < \theta < \Theta$, where $\theta$ is
single-valued, this is a legitimate quiescent state: an azimuthal gradient of activity produces the purely
radial chiral force $(\boldsymbol{\nabla}p_R)^* = (A/r)\,\hat{\boldsymbol{e}}_r$, and the fluid supports it,
at rest, through a logarithmic radial pressure profile --- the polar counterpart of
\eqref{eq:linear_example}. In the language of the ideal-flow analogy, \eqref{eq:log_state} is the
hydrostatic sibling of the point source $w = (m/2\pi)\log z$. Its companion, the vortex-like potential
$\chi = -iA\log z$, corresponds to a radial, logarithmic torque profile $\tau_a \propto \ln r$ in an
annulus: harmonic and single-valued, so that \eqref{eq:harmonic_condition} is satisfied, yet its conjugate
pressure is proportional to $\theta$ and therefore multivalued, the period
$\oint \partial_n p_R\,\mathrm{d}s = 2\pi\beta\tau_1$ failing to vanish. Despite its harmonic activity
profile, the annulus admits no rest state and the fluid is forced into azimuthal motion. The parallel with
the classical theory is exact but pointed: the ideal point vortex tolerates its multivalued potential
because only $\boldsymbol{\nabla}\phi$ is physical, and the multivaluedness merely encodes the circulation;
in the chiral fluid the pressure itself is observable, the multivaluedness is inadmissible, and the
obstruction is resolved by flowing. Finally, in the full disk, regularity at the origin excludes the
logarithm altogether: among radially symmetric activity profiles only the homogeneous state survives, and
any centred, radially structured torque distribution necessarily sets the fluid in motion. The steady
states that replace these forbidden equilibria are constructed next.

\subsection{Axisymmetric base states: forced azimuthal flow}
\label{sec:azimuthal}

The preceding analysis establishes when a chiral fluid can be at rest; we now construct the steady states that
replace the forbidden equilibria, the second exact family of base states of this work. Rest is broken here by
an inhomogeneous activity, and the flow that results is confined to the region where the activity varies; the
cavity flows of \S\ref{sec:circular} and \S\ref{sec:square}, by contrast, are driven from the boundary at
uniform activity. The two mechanisms are independent, but they share an operator, and the screening length
that emerges below is the one that governs the confined flows as well. Note first that the antisymmetric
stress $p_R\,\epsilon_{ij}$ is traceless and its tractions purely tangential: the mechanical pressure of the
fluid is $p$ alone, and rotational activity cannot renormalise the equation of state through the odd
sector. Within the present framework, therefore, no hydrostatic route to density segregation exists; the
resolution is necessarily kinetic. Consider an arbitrary axisymmetric activity profile $\tau_a(r)$ in a disk
or annulus. Unless $\tau_a$ is uniform, no quiescent state exists, and the obstruction is resolved by an
azimuthal flow $\boldsymbol{u} = u_\theta(r)\,\hat{\boldsymbol{e}}_\theta$. From the angular momentum balance
equation \eqref{eq:chiral_angular_moment} with steady flow ($\partial_t=0$) we obtain
$p_R = \beta\,\tau_a - \mu_R'\,\omega$, where $\mu_R' \equiv \mu_R\Gamma^\Omega/[2(2\mu_R+\Gamma^\Omega)]$ is
the modified rotational viscosity that will reappear in \S\ref{sec:cavity}; using the identity
$\mathrm{d}\omega/\mathrm{d}r = \mathcal{L}[u_\theta]$, with $\mathcal{L}[f] \equiv f'' + f'/r - f/r^2$, the
azimuthal momentum balance in the Stokes regime becomes

\begin{equation}
  \label{eq:azimuthal_balance}
  \left(\mu + \mu_R'\right)\mathcal{L}[u_\theta] - \Gamma\, u_\theta
  = \beta\,\frac{\partial \tau_a}{\partial r} ,
\end{equation}
a screened linear response whose homogeneous solutions are the modified Bessel functions $I_1(r/\ell)$ and
$K_1(r/\ell)$, with screening length $\ell = \sqrt{(\mu+\mu_R')/\Gamma}$. This is the same length that
governs the confined flows of \S\ref{sec:cavity}, where it reappears as $|\alpha|^{-1}$ with $\alpha$ defined
in \eqref{eq:alpha_def}. In the drag-dominated limit,
\begin{equation}
  \label{eq:edge_current}
  u_\theta = -\,\frac{\beta}{\Gamma}\,\frac{\partial\tau_a}{\partial r} ,
  \qquad
  \omega = -\,\frac{\beta}{\Gamma}\,\nabla^2\tau_a :
\end{equation}
the current is localised where the activity profile varies --- an interfacial edge current --- and the
vorticity measures, literally, the non-harmonicity of $\tau_a$ that forbade rest. The radial balance,
meanwhile, collapses onto $\partial_r p = \mu^{\mathrm{odd}}\,\partial_r\omega$, so that
\begin{equation}
  \label{eq:odd_pressure_profile}
  p(r) = \mathrm{const} + \mu^{\mathrm{odd}}\,\omega(r) ,
\end{equation}
i.e.\ the renormalised pressure $p^* = p -\mu^{\mathrm{odd}} \omega$ of \eqref{eq:pressure_shift} is uniform,
and the mechanical pressure $p$ follows the vorticity profile. Also, the mechanical pressure difference
between regions of different vorticity $\Delta p = \mu^{\mathrm{odd}}\Delta\omega$, is a pure fingerprint of
the odd viscosity --- a flow-state measurement of $\mu^{\mathrm{odd}}$ complementary to the flow-free
measurement of $\beta$ provided by the wall torque \eqref{eq:boundary_torque}. The physical scope of this base
state is broad: whenever the torque density inherits a radially segregated density field --- as in suspensions
whose transport coefficients scale with the local density --- the results above imply that the segregated
configuration cannot rest, that its interfaces necessarily carry azimuthal currents, and that its pressure
profile is set by the odd sector alone. Coexistence in a chiral fluid is, in this precise sense, not
hydrostatic but dynamical. The application of this mechanism to the rotation-induced phase separation recently
reported in chiral fluids \citep{Digregorio_2025} --- including the density dynamics and the selection of the
coexisting states, which lie beyond the constant-coefficient framework of the present work --- is developed in
a companion paper (Di~Gregorio, Pagonabarraga \& Vega~Reyes, in preparation).

\section{The chiral Stokes cavity flow}
\label{sec:cavity} 
 
Particularised to a steady state under slow flow conditions so that the advective terms
$\propto(\boldsymbol{u}\cdot\boldsymbol\nabla)$ can be neglected, the balance equations read
 
\begin{align}
  &\rho = \rho_0; \quad \nabla \cdot \boldsymbol{u} = 0 , \label{eq:stokes_mass}\\
  &\mu \nabla^2 \boldsymbol{u} + \mu^{\mathrm{odd}} \nabla^2 \boldsymbol{u^*} = \nabla p + \Gamma \boldsymbol{u} -
    \boldsymbol{\nabla}^* p_R , \label{eq:stokes_momentum}\\
  & \left(2\mu_R + \Gamma^\Omega\right)\Omega = \tau_a + \mu_R\,\omega ,\label{eq:stokes_angular}
\end{align}

If we work out the stream-vorticity formulation of \eqref{eq:stokes_momentum}--\eqref{eq:stokes_angular}
it is possible to demonstrate that these equations reduce to a modified version of the Stokes cavity flow,
with the addition of a source term $\Gamma\omega$ in the vorticity equation. For this, we need to use the
properties $\nabla\times\boldsymbol{u}^* = -(\nabla\cdot\boldsymbol{u})\boldsymbol{e}_z = 0$ (for incompressible flow) and
$\nabla\times\nabla^*\omega = - \nabla^2\omega$.

For two-dimensional incompressible flow, the continuity equation requires, as we know:
\begin{equation}
  \frac{\partial u_x}{\partial x} + \frac{\partial u_y}{\partial y} = 0.
\label{eq:continuity}
\end{equation}

Next, we make use of the stream function $\psi(x,y)$ such that \citep{Batchelor_1967}:

\begin{equation}
u_x =\frac{\partial \psi}{\partial y}, \qquad u_y = -\frac{\partial \psi}{\partial x}.
\label{eq:stream_def}
\end{equation}
Note that $\psi$ is defined only up to an additive constant, which poses no technical problem.
Substituting \eqref{eq:stream_def} into the continuity equation \eqref{eq:continuity} we obtain:

\begin{equation}
  \frac{\partial}{\partial x}\left(\frac{\partial \psi}{\partial y}\right) +
  \frac{\partial}{\partial y}\left(-\frac{\partial \psi}{\partial x}\right) =
  \frac{\partial^2 \psi}{\partial x \partial y} - \frac{\partial^2 \psi}{\partial y \partial x} = 0
\end{equation}

The mixed partial derivatives are equal (by Schwarz's theorem), and thus cancel identically. Therefore, the
continuity equation is \textit{automatically} satisfied for any possible choice of $\psi$.

From the definition of vorticity we can obtain a differential equation that couples it with the stream
function we obtain the fundamental kinematic relationship:

\begin{equation}
\omega =  \nabla \times \boldsymbol{u} = \epsilon_{ij}\partial_i u_j = \left( \frac{\partial u_y}{\partial x} -
  \frac{\partial u_x}{\partial y}\right) = \frac{\partial}{\partial x}\left(-\frac{\partial \psi}{\partial
    x}\right) - \frac{\partial}{\partial y}\left(\frac{\partial \psi}{\partial y}\right) = -\nabla^2\psi. 
\label{eq:vorticity_stream_kinematic}
\end{equation}
Note that the kinematic relationship \eqref{eq:vorticity_stream_kinematic} is valid for steady incompressible
flows for both kind I and kind II two-dimensional fluids.

We need to close the stream-vorticity description of the chiral fluid by now resorting to equations and taking
the curl of the momentum balance equation

\begin{equation}
  \mu \boldsymbol\nabla \times (\nabla^2 \boldsymbol{u}) +
  \mu^{\mathrm{odd}} \boldsymbol\nabla \times (\nabla^2 \boldsymbol{u^*}) = \boldsymbol\nabla \times
  (\boldsymbol\nabla p) + \Gamma (\boldsymbol{\nabla}\times\boldsymbol{u}) -
  \boldsymbol{\nabla}\times(\boldsymbol{\nabla}^*p_R). \label{eq:vort_trans_eq} 
\end{equation}

We note that\footnote{Two identities used below deserve an explicit check. First, the extra term
  $\boldsymbol\nabla\times(\nabla^2\boldsymbol{u^*})$ in \eqref{eq:vort_trans_eq}, due to oddness: using
  $\nabla\times\boldsymbol{u^*}\equiv\epsilon_{ij}\partial_iu_j^*$, $u^*_j\equiv\epsilon_{ji}u_i$ and
  $\epsilon_{ij}\epsilon_{ji}=-1$, one has
  $\nabla\times(\nabla^2\boldsymbol{u^*})\equiv\epsilon_{ij}\epsilon_{ji}\nabla^2\partial_iu_i=
  -\nabla^2(\boldsymbol\nabla\cdot\boldsymbol{u})=0$ for incompressible flow. Second, the curl of the rotated pressure
  gradient: with $(\boldsymbol{\nabla}^*p_R)_i\equiv\epsilon_{ij}\partial_jp_R$, the 2D scalar curl reads
  $\boldsymbol{\nabla}\times(\boldsymbol{\nabla}^*p_R)=\epsilon_{mi}\partial_m(\epsilon_{ij}\partial_jp_R)
  =\epsilon_{mi}\epsilon_{ij}\partial_m\partial_jp_R=-\delta_{mj}\partial_m\partial_jp_R=-\nabla^2p_R$, where
  we used $\epsilon_{mi}\epsilon_{ij}=-\delta_{mj}$. The rotated gradient thus behaves oppositely to the plain
  gradient: the curl of $\boldsymbol{\nabla}p$ vanishes, while that of $\boldsymbol{\nabla}^*p_R$ is minus the
  Laplacian --- this is what makes the chiral pressure a genuine source of vorticity in
  \eqref{eq:pre_vorticity_helm}.}:

\begin{itemize}
\item $\boldsymbol\nabla \times (\boldsymbol\nabla p) = \boldsymbol{0}$ (the curl of a gradient is identically zero).
\item $\boldsymbol\nabla \times (\boldsymbol\nabla^* p_R) = -\nabla^2p_R $.
\item $\boldsymbol\nabla \times \boldsymbol{u} = \omega$ (vorticity).
\item $\boldsymbol\nabla \times (\nabla^2 \boldsymbol{u}) = \nabla^2(\nabla \times \boldsymbol{u}) =
  \, \nabla^2 \boldsymbol{\omega}$ (for sufficient smoothness).
\item $\boldsymbol\nabla\times(\nabla^2\boldsymbol{u^*}) = -\nabla^2(\boldsymbol\nabla\cdot\boldsymbol{u}) = 0$.
\end{itemize}
This leads to a simplified equation

\begin{equation} 
  \label{eq:pre_vorticity_helm}
  \mu\nabla^2\omega = \Gamma \omega + \nabla^2 p_R
\end{equation}

Assuming spatially constant activity (i.e., $\tau_a = \mathrm{const.}$), which is the case for a
homogeneous fluid, and taking the Laplacian of \eqref{eq:stokes_angular}:

\begin{equation}
\nabla^2 \Omega = \frac{\mu_R}{2\mu_R + \Gamma^\Omega} \nabla^2 \omega
\end{equation}

Therefore:

\begin{equation}
  \nabla^2 p_R = \mu_R \nabla^2 \left(\Omega - \tfrac{1}{2}\omega\right) =
  \frac{\mu_R^2}{2\mu_R + \Gamma^\Omega} \nabla^2 \omega -
\frac{\mu_R}{2}\nabla^2 \omega = - \mu_R \frac{\Gamma^\Omega}{2(2\mu_R + \Gamma^\Omega)} \nabla^2 \omega, 
\end{equation}
which can be used into the vorticity equation. Before this, we define the \textit{modified rotational
  viscosity}, which as we will see is an important parameter:

\begin{equation}
\mu_R' \equiv \mu_R \frac{\Gamma^\Omega}{2(2\mu_R + \Gamma^\Omega)} 
\end{equation}

Therefore:

\begin{equation}
  (\mu + \mu_R') \nabla^2 \omega = \Gamma \omega \quad \Rightarrow \quad \nabla^2 \omega =
  \frac{\Gamma}{\mu + \mu_R'} \omega
\end{equation}

And, finally, the complete system of equations in the stream-vorticity formulation is

\begin{align}
  \nabla^2 \psi &= - \omega \quad \text{(Kinematic equation)} \label{eq:stream_vor_psi} \\
  \nabla^2 \omega &= \alpha^2\omega \quad \text{(Dynamic equation,
                    \textit{modified Helmholtz equation})}. \label{eq:stream_vor_omega}
\end{align}

where:

\begin{equation}
  \label{eq:alpha_def}
\alpha^2 \equiv \frac{\Gamma}{\mu + \mu_R'} ,
\end{equation}
so that $\alpha^{-1}$ has dimensions of length and plays the role of a Yukawa-type screening length for the
vorticity field: in the dissipative branch ($\mu+\mu_R'>0$), $\alpha$ is real and vorticity imposed at the
boundaries decays into the bulk over the distance $\alpha^{-1}$; in the active branch ($\mu+\mu_R'<0$),
$\alpha^2<0$ and the solutions become oscillatory, so that vorticity is pumped rather than screened, in exact
analogy with evanescent versus propagating waves. We denote $\alpha^{-1}$ as the \textit{chiral screening
  length}. Confined flows in a domain of characteristic size $L$ are governed by the single dimensionless
group $\alpha L$.

Therefore, the 2D chiral fluid obeys, for steady incompressible flows, a set of differential equations
analogous to the well-known Stokes cavity \citep{Shankar_2000}, with the right term in the dynamic equation
being the only difference here. As a consequence, analogous numerical methods to those employed for the cavity
incompressible flow \citep{Shankar_2000,FP02} can be used in order to solve the problem of the simplest steady
chiral flow. Henceforth, we denote this type of flow as \textit{chiral Stokes cavity flow}.

The stream-vorticity formulation has notable advantages:

\begin{enumerate}
\item Pressure elimination: The pressure variable is completely eliminated from the system.
\item Automatic continuity: The continuity equation is satisfied identically.
\item Reduced system: Only two scalar equations instead of three (two momentum + continuity).
\item Efficient numerics: both equations are linear elliptic problems (Poisson and modified Helmholtz), for
  which efficient solvers exist.
\item Physical insight: Stream function provides natural visualization (streamlines), while vorticity
  represents rotation.
\end{enumerate}

In summary, the stream function-vorticity formulation provides an elegant and efficient approach to solving
two-dimensional incompressible flows. By eliminating pressure and automatically satisfying continuity, it
reduces the complexity of the Navier--Stokes equations while maintaining clear physical interpretation. The
primary challenge lies in the proper treatment of boundary conditions for vorticity, which must be carefully
derived from the no-slip condition through the stream function.

In the case of active matter, however, the no-slip condition is usually violated, and thus this condition is
not to be taken into account; the working conditions adopted in this work are stated below in
\eqref{eq:cavity_bc}. This boundary slip feature is probably linked, in the case of active rotation, to the
emergence of chiral convective patterns.

\subsection{Physical interpretation}
\label{sec:cavity_interpretation}

The sign of $\alpha^2$ selects two qualitatively different regimes of the modified Helmholtz equation
\eqref{eq:stream_vor_omega}. When $\mu + \mu_R'>0$ --- the \textit{dissipative branch} --- $\alpha^2>0$
and vorticity imposed at the boundaries decays exponentially into the bulk over the screening length
$\alpha^{-1}$: the fluid dampens the vorticity injected by the walls, and the flow relaxes towards
irrotational behaviour far from them. When $\mu + \mu_R'<0$ --- the \textit{active branch}, admissible in
the counter-rotating scenario $\mu_R<0$ with sufficient rotational drag --- $\alpha^2<0$ and $\alpha$
becomes imaginary, so that the vorticity field acquires oscillatory (propagating) solutions rather than
decaying ones, and boundary vorticity is amplified into the bulk instead of screened. The transition
$\alpha^2=0$ recovers the classical Stokes cavity for two-dimensional incompressible flow.

Two features of the equation deserve emphasis. First, the translational and rotational drags
$\Gamma$ and $\Gamma^\Omega$ are essential to the chiral coupling: setting $\Gamma \to 0$ removes the source
term altogether ($\alpha^2\to 0$), so that the coupling of the spin field to the vorticity through the
chiral pressure vanishes in the absence of substrate friction --- the confined chiral flow is genuinely a
suspension phenomenon. Second, the modified rotational viscosity $\mu_R'$ enters the effective transport
coefficient in the same additive combination $\mu+\mu_R'$ that governs the axisymmetric base state of
\S\ref{sec:azimuthal}; the screening length $\alpha^{-1}$ found here coincides with the one identified
there, so that a single mesoscopic scale controls both one- and two-dimensional confined flows in the
theory.

\subsection{The circular cavity: closed-form solution}
\label{sec:circular}

\begin{figure}[t!]
  \centering
  \includegraphics[width=\textwidth]{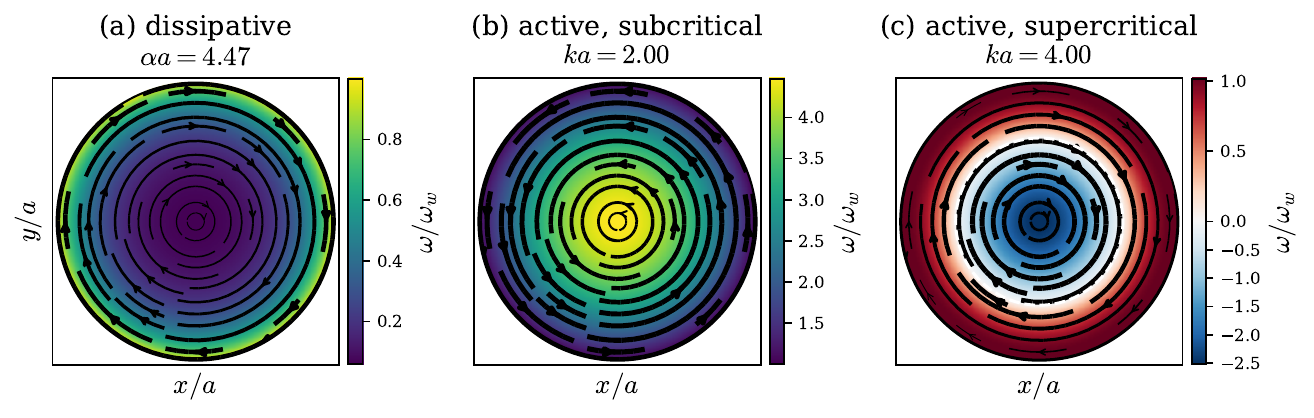}
  \caption{Chiral Stokes cavity flow in a circular domain of radius $a$, from the closed-form solution
    \eqref{eq:circ_omega}--\eqref{eq:circ_psi}. Colour: vorticity $\omega/\omega_w$; streamlines with
    thickness proportional to $|\boldsymbol{u}|^{1/2}$. (a)~Dissipative branch, $\alpha$ real:
    $\omega \propto I_0(\alpha r)$, monotonic and nodeless. (b)~Active branch with $ka<j_{0,1}=2.405$:
    $\omega \propto J_0(kr)$, amplified towards the centre but still of one sign. (c)~Active branch with
    $j_{0,1}<ka<j_{0,2}$: $J_0(kr)$ possesses one interior zero, marked by the dashed circle at
    $r/a = j_{0,1}/ka = 0.601$, across which the vorticity reverses sign; the dotted circle at
    $r/a = j_{1,1}/ka = 0.958$ marks the zero of $u_\theta$, the inner boundary of a thin cell of reversed
    circulation. Note the individual colour scale
    of each panel. Comparison with figure~\ref{fig:cavity_square} shows that the phenomenology is
    controlled by the dimensionless group $|\alpha| a$, respectively $|\alpha| L$, and not by the geometry:
    the two critical values, $j_{0,1}$ and $\pi\sqrt{2}$, are the square roots of the fundamental Dirichlet
    eigenvalues of the respective domains.}
  \label{fig:cavity_disk}
\end{figure}

The chiral Stokes cavity problem is posed by the working convention of a constant vorticity on the walls,
together with impermeability. In a domain $\mathcal{D}$ with boundary $\partial\mathcal{D}$ --- a disk of
radius $a$ or a square of side $L$ --- and uniform activity $\tau(\mathcal{A})$, the conditions are
\begin{equation}
  \label{eq:cavity_bc}
  \omega = \omega_w \quad\text{and}\quad \psi = 0 \quad\text{on } \partial\mathcal{D} ,
\end{equation}
with $\Omega$ requiring none: it is slaved pointwise to $\omega$ by the algebraic closure
\eqref{eq:stokes_angular}. Note that only one condition is imposed on $\psi$; consistently with the absence
of no-slip in active suspensions, the tangential wall velocity is an output of the solution rather than a
datum.

In a circular domain \eqref{eq:cavity_bc} becomes a statement of axisymmetry, and the system
\eqref{eq:stream_vor_psi}--\eqref{eq:stream_vor_omega} admits a closed-form solution. With
$\partial_\theta\equiv 0$, \eqref{eq:stream_vor_omega} is Bessel's equation of order zero, and the solution
regular at the origin --- $K_0$ being discarded --- is
\begin{equation}
  \label{eq:circ_omega}
  \omega(r) = \omega_w\,\frac{I_0(\alpha r)}{I_0(\alpha a)} .
\end{equation}
Since $\nabla^2 I_0(\alpha r) = \alpha^2 I_0(\alpha r)$, the kinematic equation \eqref{eq:stream_vor_psi} is
integrated by inspection, the particular solution being $-\omega/\alpha^2$ and the regular homogeneous solution a
constant fixed by $\psi(a)=0$:
\begin{equation}
  \label{eq:circ_psi}
  \psi(r) = \frac{\omega_w}{\alpha^2}\left[1 - \frac{I_0(\alpha r)}{I_0(\alpha a)}\right] , \qquad
  u_\theta(r) = -\frac{\mathrm{d}\psi}{\mathrm{d}r}
  = \frac{\omega_w}{\alpha}\,\frac{I_1(\alpha r)}{I_0(\alpha a)} , \qquad u_r = 0 ,
\end{equation}
together with $\Omega(r) = \tau(\mathcal{A})/(2\mu_R+\Gamma^\Omega) +
\mu_R\,\omega(r)/(2\mu_R+\Gamma^\Omega)$ from \eqref{eq:stokes_angular}: a uniform offset set by the
activity, superposed on the vorticity profile. Expanding $I_0(x)\simeq 1+x^2/4$ recovers rigid-body
rotation, $u_\theta \to \omega_w r/2$ with $\omega\to\omega_w$, as $\alpha\to 0$.

\begin{figure}[t!]
  \centering
  \includegraphics[width=\textwidth]{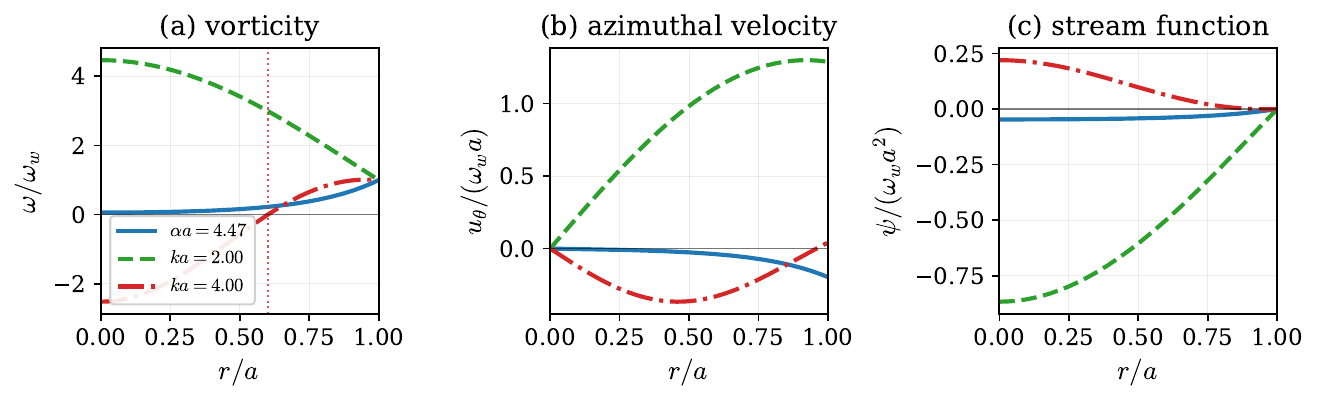}
  \caption{Radial profiles of (a) vorticity, (b) azimuthal velocity and (c) stream function in the circular
    cavity, from \eqref{eq:circ_omega}--\eqref{eq:circ_psi}, for the three regimes of
    figure~\ref{fig:cavity_disk}. In the supercritical case the vorticity and the circulation reverse at
    distinct radii: $\omega$ vanishes at $r_1 = j_{0,1}/k = 0.601\,a$ (dotted line), whereas $u_\theta$
    vanishes at $r'_1 = j_{1,1}/k = 0.958\,a$ (dash-dotted line), where $\psi$ attains an interior
    extremum and a thin cell of reversed circulation appears against the wall. In the dissipative and
    subcritical regimes all three fields are of one sign and the flow is a single vortex.}
  \label{fig:cavity_profiles}
\end{figure}

Equations \eqref{eq:circ_omega}--\eqref{eq:circ_psi} hold in both branches of
\S\ref{sec:cavity_interpretation}, which are their two analytic continuations (figures~\ref{fig:cavity_disk}
and \ref{fig:cavity_profiles}). In the dissipative branch $\alpha$ is real and $I_0$ is monotonic and
nodeless: the wall vorticity is screened over $\alpha^{-1}$ and the flow is a single vortex. In the active
branch $\alpha$ is purely imaginary; writing $\alpha=\mathrm{i}k$ with $k\equiv|\alpha|$ and using
$I_\nu(\mathrm{i}x)=\mathrm{i}^\nu J_\nu(x)$, the profile becomes $\omega(r) = \omega_w J_0(kr)/J_0(ka)$ with
$u_\theta(r) = (\omega_w/k)\,J_1(kr)/J_0(ka)$, and the distinction between the branches is exactly that
between a real and a purely imaginary wavenumber.

The oscillatory branch is singular at $J_0(ka)=0$, i.e.\ at $ka = j_{0,m}$ with $j_{0,1}=2.405$,
$j_{0,2}=5.520,\dots$ the zeros of $J_0$. The divergence signals that no steady state exists there, the
activity feeding an eigenmode of the cavity that no dissipative channel balances. Below the first such
radius the vorticity is of one sign throughout; for $j_{0,m}<ka<j_{0,m+1}$ the profile has $m$ interior
zeros at $r_i = j_{0,i}/k$, across which the vorticity reverses sign, so that the cavity is partitioned into
$m+1$ concentric annuli of alternating local rotation. The circulation possesses its own critical set,
interlaced with the first: by \eqref{eq:circ_psi}, $u_\theta \propto J_1(kr)$, which vanishes at the radii
$r'_n = j_{1,n}/k$, where $\psi$ attains an interior extremum and the flow splits into nested,
counter-rotating cells. Since $j_{0,m} < j_{1,m} < j_{0,m+1}$, the two structures alternate as $ka$ grows,
each reversal of the vorticity being followed by the birth of a flow cell. In the window
$j_{0,1}<ka<j_{1,1}$, with $j_{1,1}=3.832$, the vorticity has reversed once while the flow is still a
single vortex --- one that circulates opposite to the rigid-rotation limit, the amplitude $1/J_0(ka)$
having changed sign through the resonance; for $j_{1,1}<ka<j_{0,2}$, the case of
figure~\ref{fig:cavity_disk}(c), a thin cell of restored circulation appears against the wall, its inner
boundary at $r'_1 = j_{1,1}/k$. A vorticity node and a circulation reversal are thus distinct events at
distinct radii: at $r_1$ the local spin of fluid elements reverses within an unbroken cell, whereas at
$r'_1$ the advective circulation itself reverses (figure~\ref{fig:cavity_profiles}). Since $ka$ is the
dimensionless group of \S\ref{sec:cavity_interpretation} evaluated on the disk, these transitions may be
crossed by enlarging the cavity at fixed material parameters. None of this structure has a Newtonian
counterpart: for a fluid of the kind I $\nabla^2\omega=0$, the maximum principle applies unconditionally
and $\omega\equiv\omega_w$ throughout.

\subsection{The square cavity: numerical solution}
\label{sec:square}

In a square domain of side $L$ no analogous closed form exists, because a constant $\omega$ along four
straight walls is incompatible with a single separable mode: the corners couple the harmonics, and
\eqref{eq:stream_vor_psi}--\eqref{eq:stream_vor_omega} must be solved numerically, under the same conditions
\eqref{eq:cavity_bc}. We integrate the two Dirichlet problems successively --- the modified Helmholtz equation
for $\omega$, then the Poisson equation for $\psi$ --- by second-order finite differences on a uniform grid,
with the walls excluded from the linear system and their values entering as source terms; the velocity follows
by differentiation of $\psi$ and the spin field pointwise from \eqref{eq:stokes_angular}. Details are given in
Appendix~\ref{app:numerics}.

\begin{figure}[t!]
  \centering
  \includegraphics[width=\textwidth]{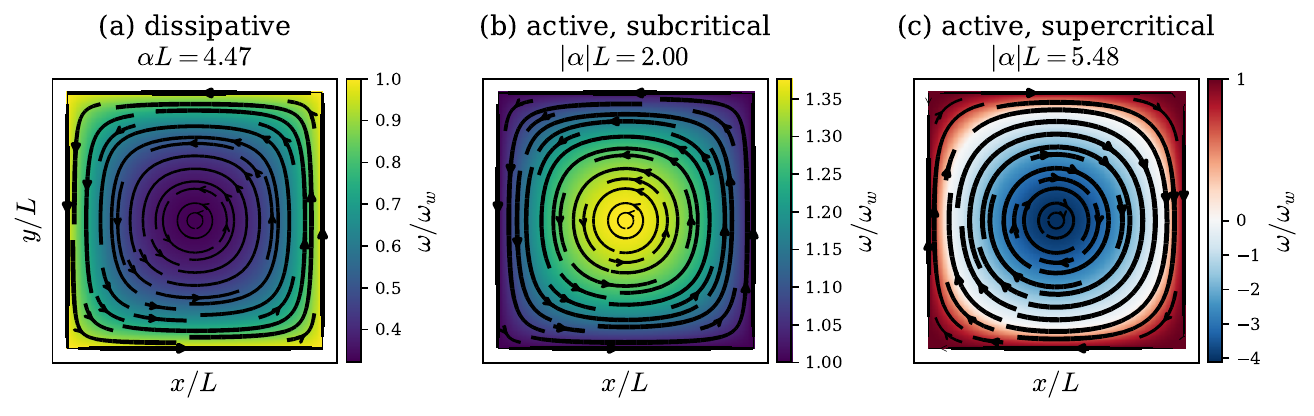}
  \caption{Chiral Stokes cavity flow in a square domain of side $L$, obtained numerically under the
    boundary conditions \eqref{eq:cavity_bc} with $\omega_w>0$. Colour: vorticity $\omega/\omega_w$;
    streamlines with thickness proportional to $|\boldsymbol{u}|^{1/2}$. (a)~Dissipative branch,
    $\alpha^2>0$: the wall vorticity is screened monotonically into the bulk over the length $\alpha^{-1}$
    and the interior relaxes towards irrotational behaviour. (b)~Active branch below the first Dirichlet
    eigenvalue, $|\alpha| L < \pi\sqrt{2}$: the vorticity is amplified towards the centre but retains a
    single sign, and the flow remains a single vortex. (c)~Active branch above it,
    $|\alpha| L > \pi\sqrt{2}$: the vorticity reverses sign in the interior and the bulk circulation is
    reversed relative to~(b). Note the individual colour
    scale of each panel.}
  \label{fig:cavity_square} 
\end{figure}

The results, shown in figure~\ref{fig:cavity_square}, reproduce the phenomenology of \S\ref{sec:circular} in
full: monotonic screening over $\alpha^{-1}$ in the dissipative branch, amplification towards the centre at
fixed sign in the subcritical active one, and interior reversal of the sign of the vorticity beyond the first
Dirichlet eigenvalue, here at $|\alpha| L = \pi\sqrt{2}$, accompanied by a reversal of the bulk circulation
between the subcritical and the supercritical panels and, in the latter, by weak cells of restored circulation
lodged in the corners, three orders of magnitude weaker than the bulk flow. The corner location of these
cells is explained by the eigenfunction series \eqref{eq:app_series} of appendix~\ref{app:numerics}: just
beyond the criticality the interior is dominated by the fundamental mode $\varphi_{11}$, reversed in sign
through the resonance and superposed on the wall-signed background $\omega_w$; since $\varphi_{11}$
vanishes quadratically towards the corners --- as the product of its two sine factors --- but only linearly
towards the edge midpoints, the background outlives the reversed mode precisely there, and the corners are
the last regions where the original sense of circulation survives. The agreement is not a coincidence
of geometry. The mechanism resides in the operator $\nabla^2-\alpha^2$, whose Dirichlet spectrum is discrete
in any bounded domain; what the circular geometry adds is that the solution itself, and not merely its
spectrum, can be written down. Both geometries are governed by the same dimensionless group --- $|\alpha| a$
and $|\alpha| L$ respectively --- and the chiral Stokes cavity is therefore a property of the theory rather
than an artefact of the circular symmetry that renders it solvable.

The python language code for all calculations in this subsection and in \S\ref{sec:circular} is available
online \citep{VegaReyes_2026}.



\section{Discussion and conclusion}
\label{sec:conclusion}

\subsection{Generality of the framework}
\label{sec:discussion}

We have derived the full mass, momentum and angular-momentum balance equations for a two-dimensional fluid
with a non-vanishing spin field, which we coin the kind II fluid, at different levels. Equations
\eqref{eq:mass_conservative_form}--\eqref{eq:angular_momentum_conservative_form} express the conservative form
whereas equations \eqref{eq:mass_convective_form}--\eqref{eq:angular_momentum_convective_form} represent the
convective form. For the latter, we have clearly identified that the angular momentum balance contains the
Hodge dual of the stress tensor. Equations
\eqref{eq:incompressible_density}--\eqref{eq:incompressible_angular_moment} are the balance equations for
incompressible flows in the kind II fluid. Finally, we also derive the hydrostatic state equations in
\eqref{eq:hydrostatic_moment}--\eqref{eq:hydrostatic_angular_moment},
\eqref{eq:hydrostatic_harmonic_p}--\eqref{eq:harmonic_condition} and the steady Stokes flow equations in
\eqref{eq:stokes_mass}--\eqref{eq:stokes_angular}. Each of these systems is exact within its stated regime
--- no more than $\boldsymbol{u}=\boldsymbol{0}$ for the quiescent states, the neglect of advection for
Stokes flow, and constant transport coefficients together with the bulk closure of \S\ref{sec:spin_flux} for
the incompressible balances --- and retains every term admissible at that level of description. They can
therefore be safely reused in other applications, and should encompass theoretical descriptions in the
literature.

As an illustration, two recent hydrodynamic descriptions of chiral active fluids are recovered as particular
cases of the present framework. \citet{Han_2021} measure a phenomenological viscosity matrix in a chiral
granular gas whose entries map one-to-one onto the coefficients derived here: their compression--rotation
viscosity coincides, up to sign conventions, with the odd bulk viscosity $\zeta^{\mathrm{odd}}$, their
four-index odd-viscosity tensor is the single rotated deviatoric channel, and their rotation--rotation
viscosity --- observed to deviate from the rotational friction --- takes here the entrainment-renormalised
form $\mu_R'$. \citet{Marconi_2026}, in turn, coarse-grain a model of particles interacting through
transverse forces into a compressible Navier--Stokes equation containing an antisymmetric stress
$\chi\,\epsilon_{ij}$, and predict odd diffusion, edge currents and a bubble-like inhomogeneous phase. In
both cases the spin field is not a dynamical variable: the former clamps $\Omega$ externally through a large
rotational drag, and the latter eliminates it altogether, the transverse force of its microscopic model
being obtained from a collisional model of granular spinners \citep{Digregorio_2025} by imposing that the
collisional torque balance the active one --- that is, by prescribing $\Omega$ rather than solving for it.

Prescribing the spin is precisely the $\Gamma^{\Omega}\to\infty$ limit of the theory developed here, taken at
fixed $\Omega_0(\boldsymbol{r}) = \lim \tau_a/\Gamma^{\Omega}$.\footnote{This reading has been confirmed by
  the authors of that work (L. Caprini, private communication).} In that limit the closure
\eqref{eq:stokes_angular} degenerates: the entrainment contribution remains finite,
$\beta\tau_a \to \mu_R\Omega_0$, so that a prescribed antisymmetric stress $\mu_R\Omega_0\,\epsilon_{ij}$
survives --- their $\chi$, with the density dependence of their coefficient playing the role of the spatial
variation of $\Omega_0$ --- while the feedback term saturates at $\mu_R' \to \mu_R/2$ and the chiral pressure
ceases to respond to the vorticity. What is lost in the reduction is therefore not the chiral stress but its
coupling to the flow: the angular momentum drained from the spin by the antisymmetric channel, and returned
to the orbital sector as the source $-\nabla^2 p_R$ of vorticity, is replaced by a source fixed from outside.
Consequently the reduced descriptions carry neither the entrainment coefficient $\beta$ nor the renormalised
viscosity $\mu_R'$, and hence no screening length $|\alpha|^{-1}$: the confined states of \S\ref{sec:cavity},
organised by $|\alpha|L$, have no counterpart within them.
 
The reduction is legitimate whenever the spin relaxes fast compared with the flow and is not itself structured
by the boundaries, and it buys what the present treatment does not attempt --- a density field with its own
equation of state, hence access to phase separation. The two levels of description --- the one developed in
\S\S\ref{sec:incompressible} and \ref{sec:cavity}, and that of \citet{Marconi_2026} --- are therefore
complementary, and they can be told apart experimentally: at uniform activity and uniform density, a
prescribed-spin description predicts no bulk chiral force, whereas here the spin responds to the vorticity
generated at the walls, and a boundary layer of width $|\alpha|^{-1}$ develops. The chiral Stokes cavity is
the simplest realisation of that difference.

A separate question is whether the description obtained here is unique. \citet{Markovich_2019} show that it
need not be: a chiral active particle idealised as a torque dipole of given strength admits at least three
inequivalent hydrodynamic descriptions, according to whether the dipole is resolved into two point torques,
two force pairs, or one of each, and these differ in whether the activity enters the linear or the angular
momentum balance. That ambiguity rests on the requirement that the net force and the net torque exerted by
each particle on the fluid vanish, which is what makes the decomposition of an internally generated dipole a
matter of choice. It does not arise here, for two independent reasons. First, the same authors note that in
strictly two dimensions the chiral direction is fixed normal to the plane, so that none of the three
decompositions --- all of which require the separation vector to be collinear with the torque --- can be
realised; the two-dimensional case is left open in their analysis. Second, in a kind II-a fluid the couple
$\tau_a$ is a prescribed material source and the substrate is an external sink of both momentum and angular
momentum, so that neither balance condition holds: the activity enters the spin balance as a body couple and
nowhere else, and there is nothing left to decompose.

\subsection{Conclusions}
\label{sec:conclusions_summary}

We have identified two subgroups of kind II fluids according to the origin of the net body couple that
sustains the spin field: the kind II-a fluid, subject to internal torques at the particle level --- the
chiral active fluid belonging to this class --- and the kind II-b fluid, where the net body couple is
produced by external fields. For kind II fluids we have derived the complete expressions of the stress
tensor and of the spin flux.

We write the stress tensor, with no approximation at Navier--Stokes order \citep{Kim_1998,Brush_1972}, as
the sum of the fully symmetric stress tensor of the Newtonian fluid, $\boldsymbol\sigma^{\mathrm{I}}$, and the
odd stress tensor $\boldsymbol\sigma^{\mathrm{odd}}$, which contains an isotropic antisymmetric part --- the
chiral pressure $p_R$ --- and a rotated deviatoric part --- the odd viscosity --- which is fully
symmetric. All odd contributions are generated from the Newtonian stress by the single translation--rotation
correspondence $\delta_{ij}\mapsto\epsilon_{ij}$, which also organises the transport-coefficient structure of
the theory: each coefficient of the even sector, $\{\zeta,\mu,\kappa\}$, acquires a non-dissipative
pseudoscalar partner, $\{\zeta^{\mathrm{odd}},\mu^{\mathrm{odd}},\kappa^{\mathrm{odd}}\}$, while the isotropic
channel maps the pressure onto the chiral pressure, introducing the rotational viscosity $\mu_R$. This set
exhausts the admissible linear response at Navier--Stokes order in the trace-free sectors of the stress, and
the structural parallelism between $\{\mu,\mu^{\mathrm{odd}}\}$ in the stress and
$\{\kappa,\kappa^{\mathrm{odd}}\}$ in the spin flux had not, to our knowledge, been emphasised before.

Much of this economy stems from formulating the constitutive theory in the deviatoric basis: because $d_{ij}$
is irreducible under rotations, its rotated image $\epsilon_{ik}d_{kj}$ is automatically symmetric, traceless
and free of compression terms, so that each odd channel is generated by a single monomial, with no
symmetrisation, no four-index tensors and no cross-couplings beyond those the rotation itself creates. The
intricacy of the chiral response is thus a property of the representation, not of the physics: one channel,
one coefficient, one mechanical action (figures~\ref{fig:stress_actions}, \ref{fig:stress_actions_normal}) ---
a structure that organises the phenomenological viscosity matrices of the chiral-fluid literature
(\S\ref{sec:discussion}).

The spin flux splits into an achiral gradient part and a chiral rotated part. At constant transport
coefficients the divergence of the chiral part vanishes identically, while the achiral part is negligible in
the bulk, its effects being confined to boundary layers of width the spin screening length $\ell_s$; the
spin flux may therefore be neglected everywhere except at boundaries.

The active torque $\tau_a$ sustains a spin field $\Omega$. Part of the injected angular momentum is
dissipated against the substrate through the rotational drag $-\Gamma^{\Omega}\Omega$; the remainder is
transferred, through the antisymmetric channel of the stress tensor, to the orbital angular momentum of the
flow. This orbital share splits in turn into a boundary contribution --- the couple exerted on the container
by the uniform part of $p_R$ --- and a bulk contribution that generates vorticity wherever the chiral
pressure is non-uniform, through source terms of the form $\nabla^2 p_R$ (\S\ref{sec:hydrostatic} and
\S\ref{sec:azimuthal}). The odd viscosity, by contrast, is dynamically silent in incompressible bulk flow
--- its force is a pure gradient, absorbed into the pressure --- so that the chiral pressure is the sole
chiral source of circulation (\S\ref{sec:odd_rotated_force}).

We have analysed the simplest incompressible steady flows of the chiral fluid. For these flows the spin
field is determined algebraically by the active torque density and, when the fluid flows, by the vorticity;
the entrainment coefficient $\beta$ and the dimensionless group $|\alpha|L$ govern the steady states. We
have described a family of quiescent states organised by a chiral complex potential and subject to a
topological existence condition; an axisymmetric state with azimuthal flow, induced by the inhomogeneities
of the chiral activity; and a confined flow reminiscent of the classical Stokes cavity
\citep{Burggraf_1966,Shankar_2000}, which we term the chiral Stokes cavity and solve in closed form in a
circular geometry and numerically in a square one. The cavity is organised by $|\alpha|L$ into two branches:
a dissipative branch, in which the wall vorticity is screened and the flow is a single vortex, and an active
branch, in which it is amplified and, beyond the first Dirichlet eigenvalue of the domain --- $j_{0,1}$ for
the disk, $\pi\sqrt{2}$ for the square --- reverses sign in the interior, the circulation reorganising into
secondary counter-rotating cells. The criticalities are spectral properties of the domain, so that the
phenomenology is controlled by $|\alpha|L$ and not by the geometry.

A quantitative validation of these results against the laboratory experiments of \citet{Lopez_Castano_2022}
is under development and will be reported separately.

The present theory fixes the structure of the odd stress and of the spin flux but leaves the transport
coefficients themselves as material parameters. These are not merely phenomenological: they have recently
been computed from kinetic theory in the dilute limit. \citet{Maire_2026} obtain, from a Boltzmann--Enskog
treatment of hard disks exchanging a tangential impulse at contact, Chapman--Enskog predictions for the odd
viscosity, the odd thermal conductivity and the odd self-diffusivity, and find that chirality generates an
antisymmetric homogeneous stress corresponding to a torque density --- the kinetic-theory counterpart of the
chiral pressure $p_R$ obtained here. \citet{Eren_2025} reach an equivalent conclusion for a dilute granular
gas of rough, inelastic particles driven by a constant torque, and \citet{Lier_2026} derive analytical
expressions for the shear and odd viscosities of hard disks undergoing chiral, momentum-conserving
collisions. The coefficients entering the present framework are therefore calculable from the microscopic
dynamics, while the structure in which they enter is fixed by angular-momentum conservation.

In summary, we provide here the essential theoretical tools and dynamical elements needed to interpret and
further explore the complex dynamics of chiral fluids, and we expect the hydrodynamic framework established
here to serve as a reference for future work on chiral fluids.

\begin{bmhead}[Acknowledgments]
  I am indebted to M. A. L\'opez-Casta\~no, A. Rodr\'iguez-Rivas, F. J. Gonz\'alez Saavedra, A. M\'arquez Seco
  and A. M\'arquez Seco for their contributions to the design and implementation of the particle tracking code
  for experiments and experimental set-up design. Also, the experimental data used in this work were taken
  by them. I also thank I. Pagonabarraga, P. di Gregorio, A. Santos and A. Rodr\'iguez-Rivas for fruitful
  discussion on the hydrodynamic theory of chiral fluids.
\end{bmhead}

\begin{bmhead}[Funding]
  I acknowledge support from Ministerio de Ciencia, Innovaci\'on y Universidades (AEI) and ERDF ``A way of
  making Europe'' through project no. MCIN/AEI/10.13039/501100011033 and fellowship PRX21/00490. I also
  acknowledge support from Junta de Extremadura through contract No. GR24077.
\end{bmhead}


\begin{appen}

  \section{Boundary conditions of the chiral Stokes cavity: physical origin and numerical
    treatment}
  \label{app:numerics}

\subsection{Physical origin of the boundary data}\label{app:bc_physics}

The reduced system \eqref{eq:stream_vor_psi}--\eqref{eq:stream_vor_omega} is coupled in one direction only:
the modified Helmholtz equation for $\omega$ is autonomous, and the stream function follows from a Poisson
problem with $\omega$ as its source. The steady problem thus consists of two second-order elliptic equations
solved in succession, each of which admits exactly one boundary datum, and the pair \eqref{eq:cavity_bc} ---
$\omega=\omega_w$ and $\psi=0$ on $\partial\mathcal{D}$ --- is a complete and non-redundant set.

It is instructive to contrast this count with the classical cavity. There the physical data are kinematic:
impermeability and no slip load \textit{both} conditions onto the stream function, $\psi=0$ and
$\partial_n\psi$ prescribed, and none remains for the vorticity, whose wall value is an implicit functional
of the interior solution; discretely, it must be reconstructed from $\psi$ by closures of Thom type,
$\omega_{\mathrm{wall}} = -2\psi_1/h^2$ for a resting wall (with an additional $-2U_0/h$ on a lid moving at
speed $U_0$, $\psi_1$ being the stream function at the first interior node and $h$ the grid spacing), and
iterated to consistency \citep{FP02}. The chiral cavity inverts the count: the wall vorticity is the datum,
and the tangential wall velocity is an output. In the disk this output is explicit,
$u_\theta(a) = (\omega_w/\alpha)\,I_1(\alpha a)/I_0(\alpha a)$ from \eqref{eq:circ_psi}: the theory predicts
a boundary slip --- the edge current characteristic of chiral fluids --- rather than assuming its absence.

The physical origin of the datum $\omega_w$ lies in a wall layer that the mesoscopic theory does not
resolve. In the bulk, the rotational flow of each spinner is cancelled by that of its neighbours; at a wall
the cancellation is broken on one side, and a net tangential current survives within a layer whose width is
set by the particle scale --- comparable to the spin screening length $\ell_s$ of \S\ref{sec:spin_flux}.
Since the hydrodynamic description holds at distances large compared with $\ell_s$, it receives the
integrated effect of this unresolved layer as an effective Dirichlet value for the vorticity, in the same
spirit in which a Navier slip length condenses unresolved molecular dynamics at a boundary. Its sign carries
the sense of the spinner--wall interaction, and its magnitude is an effective property of the wall, to be
determined by experiment. The spin field requires no independent
condition, as noted in \S\ref{sec:circular}: it is slaved to $\omega$ by the algebraic closure
\eqref{eq:stokes_angular}, any microscopic spin boundary layer being likewise subsumed into $\omega_w$.

Two structural consequences follow. First, since the four walls are materially identical and the activity is
uniform, a single scalar $\omega_w$ exhausts the boundary data; the problem being linear, all fields are
proportional to it, and the solution manifold is parameterised by the group $|\alpha|L$ (respectively
$|\alpha|a$) and the geometry alone, which is why figures~\ref{fig:cavity_disk} and \ref{fig:cavity_square}
display $\omega/\omega_w$. Secondly, the data are continuous at the corners --- in contrast with the
lid-driven cavity, whose wall velocity jumps where the lid meets the resting walls --- so that no corner
singularity is introduced by the boundary conditions and no local grid refinement is required.

\subsection{Numerical treatment in the square geometry}\label{app:square_numerics}

Let the domain be the square $(0,L)^2$, discretised by a uniform grid of $N\times N$ interior nodes of
spacing $h = L/(N+1)$, the walls carrying the known values. With $T = \operatorname{tridiag}(1,-2,1)$ and
$I$ the identity of order $N$, the five-point Laplacian is the Kronecker sum
$\Delta_h = (I\otimes T + T\otimes I)/h^2$, and the two Dirichlet problems read
\begin{equation}
  \label{eq:app_discrete}
  (\Delta_h - \alpha^2\,\mathsf{I})\,\omega_h = b_h , \qquad \Delta_h\,\psi_h = -\omega_h ,
\end{equation}
where $\mathsf{I}$ is the identity of order $N^2$ and the lifting vector $b_h$ collects the boundary data:
each appearance of a wall neighbour in a stencil contributes $-\omega_w/h^2$ to the right-hand side of the
corresponding row, while $\psi_h$ carries homogeneous data. Both matrices are symmetric and sparse and are
factorised directly (SuperLU, as interfaced by \texttt{scipy}); the one-way coupling means that the solution
consists of two linear solves in fixed order, with no iteration between $\psi$ and $\omega$ and no Thom-type
closure. Velocities follow from centred second-order differences of $\psi_h$, the wall slip being extracted
by one-sided second-order differences using $\psi = 0$ on the wall, and the spin field pointwise from
\eqref{eq:stokes_angular}.

For $\alpha^2 \ge 0$ the Helmholtz matrix is negative definite and the problem is uniquely solvable on any
grid. In the active branch, $\alpha^2 = -k^2$, solvability requires that $k^2$ avoid the Dirichlet spectrum
of $-\nabla^2$ on the square, $\lambda_{mn} = \pi^2(m^2+n^2)/L^2$; moreover, since the constant boundary
datum is symmetric about both midlines of the square, only the modes with $m$ and $n$ both odd are excited,
so that the accessible resonances are $kL = \pi\sqrt{m^2+n^2}$ with odd $m,n$ --- the first at
$\pi\sqrt{2}$, as found in \S\ref{sec:square}, the next at $\pi\sqrt{10}$ --- whereas eigenvalues of even
parity, e.g.\ $kL = \pi\sqrt{5}$, are crossed without incident. Near the resonant set the discrete matrix
inherits the ill-conditioning of the continuum operator, and computations are kept at a finite distance
from it.

The scheme admits two exact benchmarks. At $\alpha = 0$ the constant lies in the kernel of $\Delta_h$ with
matching boundary data, so the discrete solution reproduces $\omega \equiv \omega_w$ to rounding error, and
$\psi_h$ solves the classical torsion problem of the square. For $\alpha \neq 0$ the exact solution is
available as an eigenfunction series --- the infinite superposition made necessary by the corner coupling
noted in \S\ref{sec:square} --- with $\varphi_{mn} = \sin(m\pi x/L)\sin(n\pi y/L)$ and
$b_{mn} = 16/(\pi^2 mn)$:
\begin{equation}
  \label{eq:app_series}
  \psi = \omega_w \sum_{m,n\ \mathrm{odd}} \frac{b_{mn}\,\varphi_{mn}}{\lambda_{mn}+\alpha^2} , \qquad
  \omega = \omega_w \Big( 1 - \alpha^2 \sum_{m,n\ \mathrm{odd}}
  \frac{b_{mn}\,\varphi_{mn}}{\lambda_{mn}+\alpha^2} \Big) ,
\end{equation}
valid in both branches away from the resonant set. The finite-difference route is nevertheless retained as
the general method --- it extends unchanged to arbitrary boundary data and domains --- with
\eqref{eq:app_series} serving as its benchmark: measured against it, the discrete solutions converge at the
expected second order in $h$, in both fields and in both branches. The grids of \S\ref{sec:square} are
chosen so that $h$ resolves the smaller of the screening length $\alpha^{-1}$ and the wavelength $2\pi/k$,
and are refined until the fields change pointwise by less than $10^{-4}$.

\end{appen}\clearpage

\bibliographystyle{jfm}
\bibliography{chiral_stokes_cavity}

\begin{thebibliography}{57}
\expandafter\ifx\csname natexlab\endcsname\relax\def\natexlab#1{#1}\fi
\def\au#1{#1} \def\ed#1{#1} \def\yr#1{#1}\def\at#1{#1}\def\jt#1{\textit{#1}}
  \def\bt#1{#1}\def\bvol#1{\textbf{#1}} \def\vol#1{#1} \def\pg#1{#1}
  \def\publ#1{#1}\def\arxiv#1{#1}\def\org#1{#1}\def\st#1{\textit{#1}}

\bibitem[Aranson \& Tsimring(2006)]{Aranson_2006}
{\sc \au{Aranson, Igor~S.} \& \au{Tsimring, Lev~S.}} \yr{2006}  \at{Patterns
  and collective behavior in granular media: Theoretical concepts}.
  \jt{Reviews of Modern Physics}  \bvol{78}~(2),  \pg{641–692}.

\bibitem[Avron(1998)]{Avron_1998}
{\sc \au{Avron, J.~E.}} \yr{1998}  \at{Odd viscosity}.  \jt{Journal of
  Statistical Physics}  \bvol{92}~(3–4),  \pg{543–557}.

\bibitem[Banerjee {\em et~al.\/}(2017)Banerjee, Souslov, Abanov \&
  Vitelli]{Banerjee2017}
{\sc \au{Banerjee, D.}, \au{Souslov, A.}, \au{Abanov, A.~G.} \& \au{Vitelli,
  V.}} \yr{2017}  \at{Odd viscosity in chiral active fluids}.  \jt{Nature
  Communications}  \bvol{8},  \pg{1573}.

\bibitem[Batchelor(1967)]{Batchelor_1967}
{\sc \au{Batchelor, G.~K.}} \yr{1967} {\em An Introduction to Fluid
  Dynamics\/}.  \publ{Cambridge University Press}.

\bibitem[Beppu {\em et~al.\/}(2021)Beppu, Izri, Sato, Yamanishi, Sumino \&
  Maeda]{Beppu2021}
{\sc \au{Beppu, K.}, \au{Izri, Z.}, \au{Sato, T.}, \au{Yamanishi, Y.},
  \au{Sumino, Y.} \& \au{Maeda, Y.~T.}} \yr{2021}  \at{Edge current and pairing
  order transition in chiral bacterial vortices}.  \jt{Proceedings of the
  National Academy of Sciences USA}  \bvol{118}~(39),  \pg{e2107461118}.

\bibitem[Bowick {\em et~al.\/}(2022)Bowick, Fakhri, Marchetti \&
  Ramaswamy]{Bowick_2022}
{\sc \au{Bowick, Mark~J.}, \au{Fakhri, Nikta}, \au{Marchetti, M.~Cristina} \&
  \au{Ramaswamy, Sriram}} \yr{2022}  \at{Symmetry, thermodynamics, and topology
  in active matter}.  \jt{Physical Review X}  \bvol{12}~(1).

\bibitem[Boyer(1997)]{Boyer1997}
{\sc \au{Boyer, P.~D.}} \yr{1997}  \at{The {ATP} synthase --- a splendid
  molecular machine}.  \jt{Annual Review of Biochemistry}  \bvol{66},
  \pg{717--749}.

\bibitem[Braginskii(1965)]{Braginskii1965}
{\sc \au{Braginskii, S.~I.}} \yr{1965}  \at{Transport processes in a plasma}.
  \bt{In {\em Reviews of Plasma Physics\/} (ed. \ed{M.~A. Leontovich})}, ,
  \vol{vol.~1},  \pg{pp. 205--311}.  \publ{New York: Consultants Bureau},
  translated from the Russian.

\bibitem[Brey {\em et~al.\/}(1998)Brey, Dufty, Kim \& Santos]{Kim_1998}
{\sc \au{Brey, J.~Javier}, \au{Dufty, James~W.}, \au{Kim, Chang~Sub} \&
  \au{Santos, Andrés}} \yr{1998}  \at{Hydrodynamics for granular flow at low
  density}.  \jt{Physical Review E}  \bvol{58}~(4),  \pg{4638–4653}.

\bibitem[Brush(1972)]{Brush_1972}
{\sc \au{Brush, S.~G.}} \yr{1972} {\em Kinetic theory\/},  \st{International
  Series of Monographs in Natural Philosophy 42},  \vol{vol.~3}.  \publ{Oxford,
  UK: Pergamon Press}.

\bibitem[Burggraf(1966)]{Burggraf_1966}
{\sc \au{Burggraf, Odus~R.}} \yr{1966}  \at{Analytical and numerical studies of
  the structure of steady separated flows}.  \jt{J. Fluid Mech.}
  \bvol{24}~(1),  \pg{113–151}.

\bibitem[Burnett(1935)]{Burnett_1935}
{\sc \au{Burnett, D.}} \yr{1935}  \at{The distribution of velocities in a
  slightly non-uniform gas}.  \jt{Proc. London Math. Soc.}  \bvol{39},
  \pg{385--430}.

\bibitem[Caprini \& Marini Bettolo~Marconi(2025)]{Caprini_2025}
{\sc \au{Caprini, L.} \& \au{Marini Bettolo~Marconi, U.}} \yr{2025}  \at{Bubble
  phase induced by odd interactions in chiral systems}.  \jt{Journal of
  Chemical Physics}  \bvol{162},  \pg{161101}.

\bibitem[Cates \& Tailleur(2015)]{Cates_2015}
{\sc \au{Cates, Michael~E.} \& \au{Tailleur, Julien}} \yr{2015}
  \at{Motility-induced phase separation}.  \jt{Annual Review of Condensed
  Matter Physics}  \bvol{6}~(1),  \pg{219–244}.

\bibitem[Chandrasekhar(1981)]{Chandrasekhar_1981}
{\sc \au{Chandrasekhar, Subrahmanyan}} \yr{1981} {\em Hydrodynamic and
  Hydromagnetic Stability\/}.  \publ{New York: Dover Publications}, originally
  published: Oxford University Press, 1961.

\bibitem[Chapman \& Cowling(1970)]{Chapman_1970}
{\sc \au{Chapman, C.} \& \au{Cowling, T.~G.}} \yr{1970} {\em The Mathematical
  Theory of Non-Uniform Gases\/}, 3rd edn.  \publ{Cambridge University Press,
  Cambridge}.

\bibitem[Chaves {\em et~al.\/}(2006)Chaves, Rinaldi, Elborai, He \&
  Zahn]{Chaves_2006}
{\sc \au{Chaves, A.}, \au{Rinaldi, C.}, \au{Elborai, S.}, \au{He, X.} \&
  \au{Zahn, M.}} \yr{2006}  \at{Bulk flow in ferrofluids in a uniform rotating
  magnetic field}.  \jt{Phys. Rev. Lett.}  \bvol{96},  \pg{194501}.

\bibitem[Condiff \& Dahler(1964)]{Condiff_1964}
{\sc \au{Condiff, Duane~W.} \& \au{Dahler, John~S.}} \yr{1964}  \at{Fluid
  mechanical aspects of antisymmetric stress}.  \jt{The Physics of Fluids}
  \bvol{7}~(6),  \pg{842--854}.

\bibitem[Dahler \& Scriven(1961)]{Dahler_1961}
{\sc \au{Dahler, J.~S.} \& \au{Scriven, L.~E.}} \yr{1961}  \at{Angular momentum
  of continua}.  \jt{Nature}  \bvol{192},  \pg{36--37}.

\bibitem[Debets {\em et~al.\/}(2023)Debets, Löwen \& {Janssen M.
  C.}]{Debets_2023}
{\sc \au{Debets, Vincent~E.}, \au{Löwen, Hartmut} \& \au{{Janssen M. C.},
  Liesbeth}} \yr{2023}  \at{Glassy dynamics in chiral fluids}.  \jt{Physical
  Review Letters}  \bvol{130}~(5).

\bibitem[Di~Gregorio {\em et~al.\/}(2026)Di~Gregorio, Pagonabarraga \&
  Vega~Reyes]{Digregorio_2025}
{\sc \au{Di~Gregorio, P.}, \au{Pagonabarraga, I.} \& \au{Vega~Reyes, F.}}
  \yr{2026}  \at{Phase separation in a chiral active fluid of inertial
  self-spinning disks}.  \jt{Phys. Rev. Lett.}  \bvol{136},  \pg{218301},
  arXiv:2504.08533.

\bibitem[Digregorio {\em et~al.\/}(2018)Digregorio, Levis, Suma, Cugliandolo,
  Gonnella \& Pagonabarraga]{Digregorio_2018}
{\sc \au{Digregorio, Pasquale}, \au{Levis, Demian}, \au{Suma, Antonio},
  \au{Cugliandolo, Leticia~F.}, \au{Gonnella, Giuseppe} \& \au{Pagonabarraga,
  Ignacio}} \yr{2018}  \at{Full phase diagram of active brownian disks: From
  melting to motility-induced phase separation}.  \jt{Physical Review Letters}
  \bvol{121}~(9).

\bibitem[Eren {\em et~al.\/}(2025)Eren, Fruchart \& Vitelli]{Eren_2025}
{\sc \au{Eren, Ege}, \au{Fruchart, Michel} \& \au{Vitelli, Vincenzo}} \yr{2025}
  A collisional model of odd fluids: from {B}oltzmann equation to chiral
  hydrodynamics,  \arxiv{arXiv: 2508.12944}.

\bibitem[Eringen(1966)]{Eringen_1966}
{\sc \au{Eringen, A.~C.}} \yr{1966}  \at{Theory of micropolar fluids}.  \jt{J.
  Math. Mech.}  \bvol{16},  \pg{1--18}.

\bibitem[Ferziger \& Peri\'c(2002)]{FP02}
{\sc \au{Ferziger, J.~H.} \& \au{Peri\'c, M.}} \yr{2002} {\em Computational
  Methods for Fluid Dynamics\/}, 3rd edn.  \publ{Berlin: Springer}.

\bibitem[Frankel(2012)]{Frankel_2012}
{\sc \au{Frankel, Theodore}} \yr{2012} {\em The Geometry of Physics: An
  Introduction\/}, 3rd edn.  \publ{Cambridge University Press}.

\bibitem[Han {\em et~al.\/}(2021)Han, Fruchart, Scheibner, Vaikuntanathan,
  de~Pablo \& Vitelli]{Han_2021}
{\sc \au{Han, M.}, \au{Fruchart, M.}, \au{Scheibner, C.}, \au{Vaikuntanathan,
  S.}, \au{de~Pablo, J.~J.} \& \au{Vitelli, V.}} \yr{2021}  \at{Fluctuating
  hydrodynamics of chiral active fluids}.  \jt{Nat. Phys.}  \bvol{17},
  \pg{1260--1269}.

\bibitem[Klymko {\em et~al.\/}(2017)Klymko, Mandal \& Mandadapu]{Klymko_2017}
{\sc \au{Klymko, Katherine}, \au{Mandal, Dibyendu} \& \au{Mandadapu,
  Kranthi~K.}} \yr{2017}  \at{Statistical mechanics of transport processes in
  active fluids: Equations of hydrodynamics}.  \jt{The Journal of Chemical
  Physics}  \bvol{147}~(19),  \pg{194109}.

\bibitem[Lier \& Matus(2026)]{Lier_2026}
{\sc \au{Lier, Ruben} \& \au{Matus, Pawe{\l}}} \yr{2026} {C}hapman--{E}nskog
  expansion for chirally colliding disks,  \arxiv{arXiv: 2602.21367}.

\bibitem[Lingam(2015)]{Lingam2015}
{\sc \au{Lingam, Manasvi}} \yr{2015}  \at{Hall viscosity: A link between
  quantum hall systems, plasmas and liquid crystals}.  \jt{Physics Letters A}
  \bvol{379}~(22–23),  \pg{1425–1430}.

\bibitem[Lou {\em et~al.\/}(2022)Lou, Yang, Ding, Liu, Chen, Zhou, Ye,
  Podgornik \& Yang]{Lou_2022}
{\sc \au{Lou, Xin}, \au{Yang, Qing}, \au{Ding, Yu}, \au{Liu, Peng}, \au{Chen,
  Ke}, \au{Zhou, Xin}, \au{Ye, Fangfu}, \au{Podgornik, Rudolf} \& \au{Yang,
  Mingcheng}} \yr{2022}  \at{Odd viscosity-induced hall-like transport of an
  active chiral fluid}.  \jt{Proceedings of the National Academy of Sciences}
  \bvol{119}~(42).

\bibitem[López-Castaño {\em et~al.\/}(2022)López-Castaño, Márquez~Seco,
  Márquez~Seco, Rodríguez-Rivas \& Reyes]{Lopez_Castano_2022}
{\sc \au{López-Castaño, Miguel~A.}, \au{Márquez~Seco, Alejandro},
  \au{Márquez~Seco, Alicia}, \au{Rodríguez-Rivas, Álvaro} \& \au{Reyes,
  Francisco~Vega}} \yr{2022}  \at{Chirality transitions in a system of active
  flat spinners}.  \jt{Physical Review Research}  \bvol{4}~(3).

\bibitem[Maire {\em et~al.\/}(2026)Maire, Petrini, Marini Bettolo~Marconi \&
  Caprini]{Maire_2026}
{\sc \au{Maire, Rapha\"el}, \au{Petrini, Alessandro}, \au{Marini
  Bettolo~Marconi, Umberto} \& \au{Caprini, Lorenzo}} \yr{2026}  \at{Kinetic
  theory of chiral active disks: Odd transport and torque density}.  \jt{The
  Journal of Chemical Physics}  \bvol{165}~(3),  \pg{034501}, arXiv:2603.04273.

\bibitem[Marchetti {\em et~al.\/}(2013)Marchetti, Joanny, Ramaswamy, Liverpool,
  Prost, Rao \& Simha]{Marchetti_2013}
{\sc \au{Marchetti, M.~C.}, \au{Joanny, J.~F.}, \au{Ramaswamy, S.},
  \au{Liverpool, T.~B.}, \au{Prost, J.}, \au{Rao, Madan} \& \au{Simha,
  R.~Aditi}} \yr{2013}  \at{Hydrodynamics of soft active matter}.  \jt{Reviews
  of Modern Physics}  \bvol{85}~(3),  \pg{1143–1189}.

\bibitem[Marini Bettolo~Marconi {\em et~al.\/}(2026)Marini Bettolo~Marconi,
  Petrini, Maire \& Caprini]{Marconi_2026}
{\sc \au{Marini Bettolo~Marconi, Umberto}, \au{Petrini, Alessandro}, \au{Maire,
  Raphael} \& \au{Caprini, Lorenzo}} \yr{2026}  \at{Emergent hydrodynamics of
  chiral active fluids: vortices, bubbles and odd diffusion}.  \jt{New Journal
  of Physics}  \bvol{28}~(6),  \pg{064401}.

\bibitem[Markovich \& Lubensky(2021)]{Markovich2021}
{\sc \au{Markovich, Tomer} \& \au{Lubensky, Tom~C.}} \yr{2021}  \at{Odd
  viscosity in active matter: Microscopic origin and 3d effects}.  \jt{Physical
  Review Letters}  \bvol{127}~(4).

\bibitem[Markovich {\em et~al.\/}(2019)Markovich, Tjhung \&
  Cates]{Markovich_2019}
{\sc \au{Markovich, Tomer}, \au{Tjhung, Elsen} \& \au{Cates, Michael~E}}
  \yr{2019}  \at{Chiral active matter: microscopic ‘torque dipoles’ have
  more than one hydrodynamic description}.  \jt{New Journal of Physics}
  \bvol{21}~(11),  \pg{112001}.

\bibitem[Melcher \& Taylor(1969)]{Melcher_1969}
{\sc \au{Melcher, J.~R.} \& \au{Taylor, G.~I.}} \yr{1969}
  \at{Electrohydrodynamics: a review of the role of interfacial shear
  stresses}.  \jt{Annu. Rev. Fluid Mech.}  \bvol{1},  \pg{111--146}.

\bibitem[Moskowitz \& Rosensweig(1967)]{Moskowitz_1967}
{\sc \au{Moskowitz, R.} \& \au{Rosensweig, R.~E.}} \yr{1967}  \at{Nonmechanical
  torque-driven flow of a ferromagnetic fluid by an electromagnetic field}.
  \jt{Appl. Phys. Lett.}  \bvol{11},  \pg{301--303}.

\bibitem[Noji {\em et~al.\/}(1997)Noji, Yasuda, Yoshida \& Kinosita]{Noji1997}
{\sc \au{Noji, H.}, \au{Yasuda, R.}, \au{Yoshida, M.} \& \au{Kinosita, K.~Jr.}}
  \yr{1997}  \at{Direct observation of the rotation of {F$_1$}-{ATPase}}.
  \jt{Nature}  \bvol{386},  \pg{299--302}.

\bibitem[Omar {\em et~al.\/}(2021)Omar, Klymko, GrandPre \&
  Geissler]{Omar_2021}
{\sc \au{Omar, Ahmad~K.}, \au{Klymko, Katherine}, \au{GrandPre, Trevor} \&
  \au{Geissler, Phillip~L.}} \yr{2021}  \at{Phase diagram of active brownian
  spheres: Crystallization and the metastability of motility-induced phase
  separation}.  \jt{Physical Review Letters}  \bvol{126}~(18).

\bibitem[Rosensweig(1985)]{Rosensweig_1985}
{\sc \au{Rosensweig, R.~E.}} \yr{1985} {\em Ferrohydrodynamics\/}.
  \publ{Cambridge: Cambridge University Press}.

\bibitem[Rosensweig {\em et~al.\/}(1990)Rosensweig, Popplewell \&
  Johnston]{Rosensweig_1990}
{\sc \au{Rosensweig, R.~E.}, \au{Popplewell, J.} \& \au{Johnston, R.~J.}}
  \yr{1990}  \at{Magnetic fluid motion in rotating field}.  \jt{J. Magn. Magn.
  Mater.}  \bvol{85},  \pg{171--180}.

\bibitem[Saville(1997)]{Saville_1997}
{\sc \au{Saville, D.~A.}} \yr{1997}  \at{Electrohydrodynamics: the
  {T}aylor--{M}elcher leaky dielectric model}.  \jt{Annu. Rev. Fluid Mech.}
  \bvol{29},  \pg{27--64}.

\bibitem[Shankar \& Deshpande(2000)]{Shankar_2000}
{\sc \au{Shankar, P.~N.} \& \au{Deshpande, M.~D.}} \yr{2000}  \at{Fluid
  mechanics in the driven cavity}.  \jt{Annu. Rev. Fluid Mech.}  \bvol{32},
  \pg{93--136}.

\bibitem[Shliomis(1972)]{Shliomis_1972}
{\sc \au{Shliomis, M.~I.}} \yr{1972}  \at{Effective viscosity of magnetic
  suspensions}.  \jt{Sov. Phys. JETP}  \bvol{34},  \pg{1291--1294}.

\bibitem[Shliomis(2021)]{Shliomis_2021}
{\sc \au{Shliomis, M.~I.}} \yr{2021}  \at{How a rotating magnetic field causes
  ferrofluid to rotate}.  \jt{Phys. Rev. Fluids}  \bvol{6},  \pg{043701}.

\bibitem[Sokolov {\em et~al.\/}(2007)Sokolov, Aranson, Kessler \&
  Goldstein]{Sokolov2007}
{\sc \au{Sokolov, A.}, \au{Aranson, I.~S.}, \au{Kessler, J.~O.} \&
  \au{Goldstein, R.~E.}} \yr{2007}  \at{Collective behavior of terminating
  bacteria}.  \jt{Physical Review Letters}  \bvol{98},  \pg{158102}.

\bibitem[Soni {\em et~al.\/}(2019)Soni, Bililign, Magkiriadou, Sacanna,
  Bartolo, Shelley \& Irvine]{Soni_2019}
{\sc \au{Soni, V.}, \au{Bililign, E.~S.}, \au{Magkiriadou, S.}, \au{Sacanna,
  S.}, \au{Bartolo, D.}, \au{Shelley, M.~J.} \& \au{Irvine, W. T.~M.}}
  \yr{2019}  \at{The odd free surface flows of a colloidal chiral fluid}.
  \jt{Nat. Phys.}  \bvol{15},  \pg{1188--1194}.

\bibitem[Sweeney \& Houdusse(2010)]{Sweeney2010}
{\sc \au{Sweeney, H.~L.} \& \au{Houdusse, A.}} \yr{2010}  \at{Structural and
  functional insights into the myosin motor mechanism}.  \jt{Annual Review of
  Biophysics}  \bvol{39},  \pg{539--557}.

\bibitem[Tsai {\em et~al.\/}(2005)Tsai, Ye, Rodriguez, Gollub \&
  Lubensky]{Tsai_2005}
{\sc \au{Tsai, J.-C.}, \au{Ye, Fangfu}, \au{Rodriguez, Juan}, \au{Gollub,
  J.~P.} \& \au{Lubensky, T.~C.}} \yr{2005}  \at{A chiral granular gas}.
  \jt{Physical Review Letters}  \bvol{94}~(21).

\bibitem[Vale {\em et~al.\/}(1985)Vale, Reese \& Sheetz]{Vale1985}
{\sc \au{Vale, R.~D.}, \au{Reese, T.~S.} \& \au{Sheetz, M.~P.}} \yr{1985}
  \at{Identification of a novel force-generating protein, kinesin, involved in
  microtubule-based motility}.  \jt{Cell}  \bvol{42},  \pg{39--50}.

\bibitem[Vega \& Pérez(2002)]{VegaReyes_2002}
{\sc \au{Vega, F.} \& \au{Pérez, A.~T.}} \yr{2002}  \at{Instability in a
  non-ohmic/ohmic fluid interface under a perpendicular electric field and
  unipolar injection}.  \jt{Physics of Fluids}  \bvol{14}~(8),
  \pg{2738–2751}.

\bibitem[{Vega Reyes}(2026)]{VegaReyes_2026}
{\sc \au{{Vega Reyes}, Francisco}} \yr{2026} {The Chiral Stokes Cavity}.
  \url{https://github.com/fvegar/chiral_stokes_cavity}.

\bibitem[Vega~Reyes {\em et~al.\/}(2014)Vega~Reyes, Santos \&
  Kremer]{Reyes_2014}
{\sc \au{Vega~Reyes, Francisco}, \au{Santos, Andrés} \& \au{Kremer,
  Gilberto~M.}} \yr{2014}  \at{Role of roughness on the hydrodynamic
  homogeneous base state of inelastic spheres}.  \jt{Physical Review E}
  \bvol{89}~(2).

\bibitem[Yoshida {\em et~al.\/}(2001)Yoshida, Muneyuki \&
  Hisabori]{Yoshida_2001}
{\sc \au{Yoshida, Masasuke}, \au{Muneyuki, Eiro} \& \au{Hisabori, Toru}}
  \yr{2001}  \at{Atp synthase — a marvellous rotary engine of the cell}.
  \jt{Nature Reviews Molecular Cell Biology}  \bvol{2}~(9),  \pg{669–677}.

\bibitem[Zaitsev \& Shliomis(1969)]{Zaitsev_1969}
{\sc \au{Zaitsev, V.~M.} \& \au{Shliomis, M.~I.}} \yr{1969}  \at{Entrainment of
  ferromagnetic suspension by a rotating field}.  \jt{J. Appl. Mech. Tech.
  Phys.}  \bvol{10},  \pg{696--700}.

\end{thebibliography}

\end{document}